\documentclass[a4paper,10pt]{article}

\usepackage{jcappub}

\usepackage{graphicx,natbib}
\usepackage[figuresright]{rotating}
\usepackage{amssymb}
\usepackage{amsmath}
\usepackage{bm}
\usepackage{ulem}

\newcommand{\App}[1]{appendix~\ref{#1}}
\newcommand{\Sec}[1]{section~\ref{#1}}
\newcommand{\Eq}[1]{equation~(\ref{#1})}

\newcommand{\nab}{\mbox{\boldmath $\nabla$} {}}

\newcommand{\be}{\begin{equation}}
\newcommand{\ee}{\end{equation}}
\newcommand{\gsim}{\gtrsim}
\newcommand{\lsim}{\lesssim}
\newcommand{\bea}{\begin{eqnarray}}
\newcommand{\eea}{\end{eqnarray}}
\newcommand{\bean}{\begin{eqnarray*}}
\newcommand{\eean}{\end{eqnarray*}}

\def\Lu{{\rm Lu}}
\def\urms{u_{\rm rms}}

\def\epsM{\epsilon_{\rm M}}
\def\epsK{\epsilon_{\rm K}}

\def\xiM{\xi_{\rm M}}

\def\kM{k_{\rm M}}
\def\kMD{k_{\rm MD}}
\def\kWT{k_{\rm WT}}

\newcommand{\SSSS}{\mbox{\boldmath ${\sf S}$} {}}

\newcommand{\BB}{{\bf{B}}}
\newcommand{\JJ}{{\bf{J}}}
\newcommand{\uu}{{\bf{u}}}

\def\half{{\textstyle{1\over2}}}
\def\onethird{{\textstyle{1\over3}}}

\def\tEEM{\tilde{\cal E}_{\rm M}}

\def\EEM{{\cal E}_{\rm M}}

\def\HHM{{\cal H}_{\rm M}}

\def\EM{E_{\rm M}}
\def\tEK{\tilde{E}_{\rm K}}
\def\tEM{\tilde{E}_{\rm M}}

\def\HM{H_{\rm M}}

\newcommand{\Fig}[1]{figure~\ref{#1}}

\def\ds{\displaystyle}

\newcommand{\yanf}[3]{, Ann. Rev. Fluid Dyn. {\bf #2}, #3 (#1).}
\newcommand{\yapj}[3]{, Astrophys.\ J.\ {\bf #2}, #3 (#1).}

\newcommand{\ymn}[3]{, Mon.\ Not.\ Roy.\ Astron.\ Soc.\ {\bf #2}, #3 (#1).}

\newcommand{\ybook}[3]{ #1, {#2} (#3)}

\newcommand{\EQ}{\begin{equation}}
\newcommand{\EN}{\end{equation}}
\newcommand{\ba}{\begin{eqnarray}}
\newcommand{\ea}{\end{eqnarray}}

\def\vA{v_{\rm A}}

\def\xiM{\xi_{\rm M}}
\def\HHM{{\cal H}_{\rm M}}

\graphicspath{{./fig/}{./png/}}

\begin{document}

\title{Statistical Properties of Scale-Invariant Helical Magnetic Fields and Applications to Cosmology}

\author[a,b,c,d,e]{Axel Brandenburg}
\author[f]{Ruth~Durrer}
\author[c,g,h]{Tina Kahniashvili}
\author[c]{Sayan Mandal\footnote{Corresponding author: Authors are listed alphabetically}}
\author[i,c]{Weichen Winston Yin}

\affiliation[a]{Laboratory for Atmospheric and Space Physics, University of Colorado, Boulder, CO 80303, USA}

\affiliation[b]{JILA and Department of Astrophysical and Planetary Sciences, University of Colorado, Boulder, CO 80303, USA}

\affiliation[c]{McWilliams Center for
Cosmology and Department of Physics, Carnegie Mellon University,
5000 Forbes Ave, Pittsburgh, PA 15213, USA}

\affiliation[d]{Nordita, KTH Royal Institute of Technology and Stockholm University,
Roslagstullsbacken 23, 10691 Stockholm, Sweden}

\affiliation[e]{Department of Astronomy, AlbaNova University Center,
Stockholm University, 10691 Stockholm, Sweden}

\affiliation[f]{D\'epartement de Physique
Th\'eorique and Center for Astroparticle Physics, Universit\'e de
Geneve, Quai E. Ansermet 24, 1211 Gen\'eve 4, Switzerland}

\affiliation[g]{Department of Physics, Laurentian University, Ramsey
Lake Road, Sudbury, ON P3E 2C, Canada}

\affiliation[h]{Abastumani Astrophysical Observatory, Ilia State University,
3-5 Cholokashvili St., 0194 Tbilisi, Georgia}

\affiliation[i]{Department of Physics, University of California Berkeley, Berkeley, CA 94720}

\emailAdd{sayanm@cmu.edu}

\abstract{We investigate the statistical properties of isotropic, stochastic,
Gaussian distributed, helical magnetic fields characterized by different
shapes of the energy spectra at large length scales and study the
associated realizability condition. We discuss smoothed magnetic fields that are commonly used when the
primordial magnetic field is constrained by observational data. We are particularly interested in scale-invariant magnetic fields
that can be generated during the inflationary stage by
quantum fluctuations.
We determine the correlation length of such magnetic fields and relate it to the infrared cutoff of perturbations produced during inflation.
We show that this scale determines the observational signatures
of the inflationary magnetic fields on the cosmic microwave background.
At smaller scales, the scale-invariant spectrum changes
with time. It becomes a steeper weak-turbulence spectrum at
progressively larger scales.
We show numerically that the critical length scale where this happens
is the turbulent-diffusive scale, which increases with
the square root of time.
}

\maketitle

\section{Introduction}

The origin of cosmic magnetic fields is one of the big
open questions in astrophysics and space physics
\cite{Widrow:2002ud,Durrer:2013pga,Subramanian:2015lua}.
It is generally thought that these magnetic fields are the
result of the amplification of weak initial seed fields.
It is also clear now that $\mu$G-strength magnetic fields were
already present in spiral galaxies (like our Milky Way) when
the universe was about a third of its present age
\cite{Vallee,Bernet:2008qp,Kronberg:2007dy}.
This poses strong constraints on the initial seed magnetic field strength
and its amplification timescale. There are two basic magnetogenesis scenarios currently under discussion:
a bottom-up (astrophysical) scenario, where the seed is typically
very weak and magnetic fields are transferred from local sources within
galaxies to larger scales \cite{Kulsrud:2007an}, and a top-down
(primordial) scenario where a significant seed field is generated
prior to galaxy formation in the early universe on scales that are now
large \cite{Kandus:2010nw}.
The primordial magnetogenesis scenario is supported by recent
observations suggesting that lower bounds of the order of
$10^{-18}$ to $10^{-19}$\,G exist for the intergalactic magnetic
fields\footnote{Initially, the $10^{-15}$--$10^{-16}$\,G limit
had been obtained \cite{Neronov:1900zz,Tavecchio:2010ja} based on
studying blazar TeV photons which produce a cascade flux in the GeV
band after absorption by the extragalactic background light (EBL).
Considering the expected cascade flux with the assumption of a constant
TeV flux gives this estimate.
These bounds have been subsequently reconsidered after accounting
for the fact that  the source observation period (of the order
of a few years) limits the flux activity in the TeV blazars
\cite{Dermer:2010mm}.
The simultaneous observations of blazars in the
GeV and TeV bands lead to weaker limits of the order of
$10^{-18}$--$10^{-19}$\,G \cite{Dermer:2010mm,Taylor:2011bn}.}, (see
refs.~\cite{Neronov:1900zz,Tavecchio:2010ja,Essey:2010nd,Taylor:2011bn,Huan:2011kp,
Vovk:2011aa,Dolag:2010ni,Takahashi:2011ac,Dermer:2010mm,Finke:2015ona},
also ref.~\cite{Arlen:2012iy} for discussions on possible
uncertainties in the measurements of blazar spectra).  In addition
to these lower limits from observations, there exist also {\it upper}
limits of the order of a few nG for the intergalactic magnetic field
\cite{Durrer:2013pga,Subramanian:2015lua}.

A cosmological magnetic field contributes to the radiation-like energy
density, and sources all three helicities of linear gravitational perturbations
\cite{Tsagas:1999ft}\footnote{These are (i) the scalar mode -- density
perturbations, (ii) the vector mode -- vorticity perturbations, and (iii)
the tensor mode -- gravitational waves, that do not have an analogy within
Newtonian physics, while Newtonian physics admits the analogy for density
(and vorticity) perturbations as magnetosonic (and Alfv\'en) waves.}
which lead to corresponding temperature and polarization anisotropies of
the cosmic microwave background (CMB).
In addition it induces  Faraday rotation of the CMB polarization
direction, and affects large-scale structure (LSS) formation;
see ref.~\cite{Subramanian:2015lua} and references therein.
All these effects can be used to constrain the magnetic field strength,
and as we show below, the magnetogenesis scenarios.
Upper limits for cosmological magnetic fields can also be obtained through CMB constraints on Faraday rotation \cite{Ade:2015cao},
and these limits are independent of those from magnetic helicity
\cite{Kahniashvili:2004gq,Campanelli:2004pm,Kosowsky:2004zh}.
In addition, upper limits on extragalactic magnetic fields can be
derived from Faraday rotation measurements of polarized emission of
distant quasars \cite{a1,a11,a2,a3}.
Other tests leading to {\it upper} limits on large-scale
correlated magnetic fields are based on their effects on big bang
nucleosynthesis (BBN) \cite{Yamazaki:2012jd}, the CMB (including
CMB fluctuations, polarization, distortions, non-gaussianity, etc,
see ref.~\cite{Ade:2015cva} and references therein), or LSS formation
(for a recent review, see \cite{Subramanian:2015lua}).
The lower limit on the intergalactic magnetic field in voids,
of the order of $10^{-18}$ G on 1\,Mpc scales, is a relatively
recent constraint in modern astrophysics
(see ref.~\cite{Miniati:2010ne}), and could very well be the result of
the amplification of a primordial cosmological field \cite{Dolag:2010ni}.

One of several plausible mechanisms for the origin of these cosmic
magnetic fields is to assume that a seed magnetic field has been
generated in the early universe \cite{Subramanian:2015lua}.
Below we discuss two basic possibilities for primordial magnetogenesis
-- inflation and cosmological phase-transitions. The purpose of this
paper is to discuss magnetic energy and helicity spectra produced by
these mechanisms and to investigate relevant length scales, which is no
longer straightforward in inflationary magnetogenesis.

{\it Inflationary Magnetogenesis}:
Magnetogenesis can occur during inflation by the amplification of
quantum vacuum fluctuations, as was shown in several
pioneering works \cite{Turner:1987bw,Ratra:1991bn}.
The rapid exponential growth and the induced stretching of the field
during inflation can produce a very large correlation length of the
observed magnetic fields today.
In addition, inflation also provides a natural way to generate modes
from quantum fluctuations of the field inside the Hubble radius,
which are subsequently converted into classical fluctuations as they
exit the Hubble horizon.
Such fields can have a scale-invariant (or a nearly scale-invariant)
spectrum. These, among several other properties, make inflationary magnetogenesis an
attractive scenario (see~\cite{Subramanian:2015lua} for a recent review).

The Maxwell action describing the electromagnetic field is conformally
invariant, and the Friedmann-Lema\^itre-Robertson-Walker (FLRW) metric
describing the evolution of the universe is conformally flat.
Therefore, the process leading to quantum excitation of the magnetic field
must break conformal invariance.
For this, one has to introduce couplings of the electromagnetic
field that break conformal invariance.
There are several possibilities to achieve this.
These include coupling of the field to the inflaton,
or to the curvature (the Riemann tensor). Several authors \cite{Turner:1987bw,Ratra:1991bn,Dolgov93,Kandus:2010nw,
Bamba:2012mi,Jimenez:2011uia,Membiela:2010rv,Membiela:2012ju,Fujita:2012rb,
Motta:2012rn,Bonvin:2011dt,Byrnes:2011aa,Campanelli:2008kh,
Das:2010ywa,Durrer:2010mq,Jimenez:2010hu,Bamba:2011si,Martin:2007ue,
Caldwell:2011ra,Campanelli:2013mea,Jain:2012jy,Kunze2010,Campanelli2008,
Campanelli2009,Lambiase2008,Lemoine1995,
Gasperini1995,Bamba2004,FujitaEtAl:2015,Atmjeet14,Sorbo2011a,Sorbo2014a,
Adshead:2015pva,Adshead:2016iae}
have explored a range of models dealing with inflationary
magnetogenesis.
The inflation-generated magnetic field scenarios should
be considered with some caution due to the possibility
of a ``strong coupling problem'' and significant backreaction
\cite{Demozzi:2009fu,Kanno:2009ei,Urban:2011bu,Demozzi2012,Sharma:2017eps,Sharma:2018kgs}, which
is not a problem for the phenomenological, effective classical model,
see ref.~\cite{Tasinato:2014fia}, for Lorentz-violating magnetogenesis
\cite{Campanelli:2015jfa},
or if the function that couples the inflaton to the electromagnetic field
has sharp and non-monotonic features \cite{Ferreira:2013sqa,Ferreira:2014hma}.

{\it Phase Transition Magnetogenesis}:
Some magnetogenesis mechanisms make use of the symmetry breaking
during cosmological phase transitions (e.g.\ electroweak or QCD)
\cite{Harrison1970,Vachaspati1991,Cornwall1997,Quashnock:1988vs,
Giovannini:1997eg,Field:1998hi,Grasso:1997nx,Stevens:2012zz,
Boyarsky2012,Urban2010,Sigl1997,Joyce1997,Henley2010,Stevens2008,
Diaz2008,Ahonen1998,Giovannini2000,Dolgov2001,Vachaspati2001,
Grasso2002,Boyanovsky2003,Boyanovsky2003b,Sigl:2002kt,
Hindmarsh1998,Enqvist1998,Semikoz:2004rr,Semikoz:2004rr2}.
If the magnetic field originated during a cosmological phase transitions,
its spectrum is constrained by the causality requirement
\cite{Durrer:2003ja}, in particular the magnetic field correlation
length must be less than or equal to the Hubble length scale at the
moment of generation.

Usually, a first-order phase transition is needed for magnetic fields
of substantial strength and correlation scales to arise, the idea being
that bubbles of the new phase start nucleating in the space filled with
the old phase; these bubbles then expand and collide with each other,
ultimately filling the entire space with the new phase.
Such processes are highly out of equilibrium (and violent),
and can generate significant turbulence, amplifying the fields
\cite{Witten:1984rs}.
In this scenario, the correlation length of the magnetic field (i.e., the
magnetic domain length scale) can be associated with the phase transition
bubble size \cite{Hogan:1983zz}.
One can also associate processes like baryogenesis with these phase
transitions \cite{Vachaspati2001}.

There are two phase transitions of interest in the early
universe -- the electroweak phase transition  occurring
at a temperature of $T\sim100\,\mathrm{GeV}$, and
the QCD phase transition occurring at $T\sim150\,\mathrm{MeV}$
\cite{Subramanian:2015lua,Widrow:2002ud,Boyanovsky2006,Durrer:2013pga}.
However, these transitions are usually (i.e., within the standard
model of particle physics) not of  first-order, but are simple
\textit{crossovers}, so the transition occurs smoothly
\cite{Aoki2006,Kajantie1996,Csikor1998}.
There has been much research about the conditions or extensions of the
standard model under which these transitions become first-order
(possibilities include having a large leptonic chemical potential for
the QCD transition \cite{Stuke09}, supersymmetric extensions for electroweak phase transitions \cite{Gorbunov11}, etc).

One of the most important properties of the primordial magnetic
fields is their helicity.
Magnetogenesis mechanisms that involve a parity
violation can lead to magnetic fields with non-zero helicity (see for example refs.~\cite{Cornwall1997,
Giovannini:1997eg,Field:1998hi,Vachaspati2001,Tashiro:2012mf,
Sigl:2002kt,Subramanian:2004uf,Campanelli:2005ye,Semikoz:2004rr,Semikoz:2004rr2,
Diaz2008,Campanelli2008,Durrer:2010mq,Sorbo2011a,Sorbo2014a,Campanelli:2015jfa,Sriramkumar:2015yza}).
It is a well known fact that magnetic helicity is an important factor defining the evolution of turbulent magnetic fields in the early universe.
In particular, helicity conservation sets constraints on the decay of magnetic
fields in the early universe leading to an inverse cascade of energy in
the helical fields.
Thus, helicity leads to magnetic fields with a larger correlation length
that decay slower compared with non-helical fields \cite{Kahniashvili:2012uj,Copi08,Brandenburg:2014mwa}.
However, in the case of nearly scale-invariant magnetic fields, the correlation
length is almost frozen in \cite{Kahniashvili:2016bkp}.

The mean magnetic energy and helicity densities are related
by the realizability condition -- one of the topics of the present study
(see section~\ref{sec:HelRealCond}).
The realizability condition limits the maximal helicity that random magnetic fields can sustain \cite{Mof78}.
Hence, magnetic fields with a lower value of helicity can be defined as
states with fractional helicity.
On the other hand, numerical simulations show that magnetic fields with
zero helicity can still undergo a slower non-helical inverse transfer
of magnetic fields \cite{Durrer:2013pga,Brandenburg:2016odr}.

We study the evolution of magnetic fields in the expanding universe
by solving the magnetohydrodynamic (MHD) evolution equations for these
fields and the fluid density and velocity.
These equations describe the {coupling between random magnetic
fields and turbulent motions,} as well as amplification
and damping.
To account for the expansion of the universe, these equations
are rewritten in terms of comoving quantities \cite{beo},
The numerical simulations are done using the {\sc Pencil Code}
(\texttt{https://github.com/pencil-code}) \cite{db}, which is a high
order finite-difference code for solving equations involving compressible
magnetohydrodynamic flows.

As mentioned above, we study the statistical properties of random helical
magnetic fields generated during the early universe, with special emphasis
on the realizability conditions and its cosmological applications.
In section \ref{sec:modeling}, we review in some detail the spectral and
statistical properties of magnetic fields, taking special care to relate
the average quantities and the various characteristic length scales to
the magnetic field spectra, and define the smoothed magnitudes of the
magnetic field and the helicity, that are widely used when analyzing
observational data.
In section~\ref{sec:HelRealCond}, we discuss the realizability condition
for generic magnetic fields, and relate it to the smoothed magnitudes.
In section~\ref{sec:DiffShapes}, we consider two different types of shape
for the energy spectrum, and formulate a method to make the realizability
condition hold consistently at all scales also for inflationary, nearly scale-invariant
fields.
In section~\ref{sec:Numerical}, we present the results of numerical
simulations, such as the evolution of the magnetic mean energy density,
the correlation length, and the fractional helicity, which grows owing to magnetic energy decay.

One of the major points of the present paper is the study of
cosmological applications.
In section~\ref{sec:Cosmology} we discuss cosmological magnetic field
amplitudes and helicity limits obtained from CMB and LSS data.
Readers familiar with MHD might want to skip over sections
\ref{sec:modeling} and \ref{sec:HelRealCond} since much of what
we present there can be found in books.

In this paper, we propose to extend
the results from the Planck satellite \cite{Ade:2015cva}, taking into account the magnetohydrodynamic
evolution of the primordial plasma until the epoch of reionization.
To the best of our knowledge, such
an analysis has not been done before.
This is important since fields with partial initial helicity become fully
helical at a later stage in their evolution \cite{Brandenburg:2017neh}.
Furthermore there is some confusion in the literature regarding the dependence
of scalar quantities like the total magnetic energy density and the rms
density on helicity. This arises from the fact that the
expression for, e.g., the spectrum of the magnetic
energy density---a fourth order correlation function in the magnetic
field---includes the helicity power spectrum. We have explicitly shown that,
after integration, the helicity drops out and does not contribute to the energy density power spectrum.
This has not previously been demonstrated, it seems.
Similar integrals involving the Lorentz force, for example,
can still remain finite, however.
Further investigation is needed, which is beyond the scope of this paper.
We show that the best way to constrain the helicity is to use parity odd CMB spectra
like $C^{TB}_\ell$ or $C^{EB}_\ell$, \cite{Caprini:2003vc,Kahniashvili:2005xe}
that are zero if there is no magnetic helicity (and/or other parity violating other sources).
We revisit the current upper limits of ref.~\cite{Ade:2015cva} on the magnetic
field, accounting for the magnetic field coupling with the primordial
plasma, and consequently MHD turbulence evolution from the moment of
generation until the recombination epoch.
We also obtain the upper bounds on maximal inflationary, nearly scale-invariant
magnetic helicity accounting for  magnetic field evolution
\cite{Kahniashvili:2016bkp} and the combined upper limits on the magnetic
field strength (through CMB and LSS data) \cite{Kahniashvili:2012dy}.
In section~\ref{sec:Concl}, we present some concluding remarks.

Throughout the paper, we set $\hbar=c=k_B=1$, and we use the
Lorentz-Heaviside units to express magnetic fields, such that the magnetic
energy density is $\rho_{\rm M}({\bf x})=\mathbf{B}^2({\bf x})/2$.
Unless otherwise specified, we imply a summation over repeated indices,
and Latin indices run through $1,\dots,3$.

\section{Modeling a helical magnetic field}
\label{sec:modeling}

Seed magnetic fields are generated in the early universe (see
ref.~\cite{Subramanian:2015lua} for a review of possible primordial
magnetogenesis scenarios) from either random quantum
fluctuations during inflation, or from a first order phase transition,
which proceeds via bubble nucleation, a violent and stochastic process.
It is not surprising, therefore, that the generated magnetic fields
are themselves random and stochastic. As already mentioned, the considerations presented in this section are
not new, but in order to eliminate recent confusions in the literature,
we want to lay them out carefully and clearly.

To define correlation functions or power spectra, we take an ensemble
average, i.e., an
average over many realizations. We assume that the generated magnetic field is statistically homogeneous
and isotropic and that it obeys Gaussian statistics.

\subsection{Magnetic field spectrum}

All statistical information of a stochastically homogeneous and isotropic
Gaussian magnetic field can be obtained through its two-point correlation
function, $\mathcal B_{ij} ({\bf r}) \equiv \langle B_i(\mathbf x) B_j(\mathbf x+\mathbf r) \rangle$ \cite{monin-yaglom}, which, in its most
general form, can be written as
\begin{equation}
{\mathcal B}_{ij}({\bf r}) = M_{\rm N}(r)\delta_{ij} +\big[ M_{\rm L}(r)-M_{\rm N}(r)\big ]
{\hat r}_i{\hat r}_j+M_{\rm H}(r)\epsilon_{ijl} r_l,
\label{correlreal}
\end{equation}
where $\langle\cdots\rangle$ denotes the average over the statistical
ensemble and ${\hat r}_i=r_i/|{\bf r}|$.
In this case, it is equivalent to the volume average over all $\mathbf x$
due to homogeneity.
The functions $M_{\rm N}(r)$, $M_{\rm L} (r)$, and $M_{\rm H}(r)$ are the \textit{lateral}
(normal, N), \textit{longitudinal} (L), and \textit{helical} (antisymmetric, H)
components of the magnetic field correlation function, respectively.
Due to isotropy, all components depend only on $r\equiv |{\bf r}|$
(as there is no a preferred direction, ~$\langle {\bf B}({\bf x})
\rangle =0$, for a stochastic magnetic field).
Since $\varepsilon_{ijk}$ is invariant under rotation, the rotational
symmetry is preserved also for antisymmetric helical fields.

Although ${\mathcal B}_{ij}({\bf r})$ is rotationally invariant, the presence of the antisymmetric part ($\propto \varepsilon_{ijl} r_l M_{\rm H}(r)$) means that parity (mirror) symmetry is violated. It is easy to see that
\begin{equation}\label{parity}
{\mathcal B}_{ij}({\bf r}) = {\mathcal B}_{ji}(-{\bf r}) \,.
\end{equation}
The contribution of the helical part to each diagonal term ${\mathcal B}_{ii}$ (no summation here) of the magnetic field two-point
correlation function vanishes.
Thus, the diagonal terms, and hence the trace, {\it do not} contain
any information on the asymmetric part. This statement is true for both
solenoidal (divergence-free) and irrotational (curl-free) fields.

The normal, $M_{\rm N}(r)$, longitudinal, $M_{\rm L}(r)$, and helical, $M_{\rm H}(r)$ components of the magnetic correlation function are obtained
from the correlation function  ${\mathcal B}_{ji}$ via the following
projection operations:
\begin{eqnarray}
P_{ij}({\bf {\hat r}}){\mathcal B}_{ij} ({\bf r}) &= &2M_{\rm N}(r),
\label{M-N}
\\
{\hat r}_i {\hat r}_j {\mathcal B}_{ij} ({\bf r}) &=& M_{\rm L}(r),
\label{M-L}
\\
\varepsilon_{ijm} {\hat r}_m {\mathcal B}_{ij} ({\bf r}) &= & 2r M_{\rm H}(r),
\label{M-H}
\end{eqnarray}
where $P_{ij}({\bf {\hat r}}) = \delta_{ij} - {\hat r}_i {\hat r}_j$
is the projector tensor into the plane normal to $\bf r$. Indeed,
the trace, given by ${\mathcal B}_{ii}({\bf r}) = \delta_{ij}{\mathcal
B}_{ij} ({\bf r}) =2M_{\rm N}(r) +M_{\rm L}(r)$, is independent of the
antisymmetric part.

Since the magnetic field is divergence-free,
$\nabla\cdot\mathbf B = 0$, $M_{\rm L}$ and $M_{\rm N}$
are not independent but related by
\begin{equation}
M_{\rm N}(r) = \frac{1}{2r}\frac{d}{dr}\Big [ r^2M_{\rm L}(r) \Big]=M_{\rm L}(r) + \frac{r}{2} \frac{d}{dr} M_{\rm L}(r).
\label{div-condition}
\end{equation}
Hence, there are only two independent functions $M_{\rm L}(r)$ and $M_{\rm H}(r)$
that determine the full magnetic two-point correlation function.
Requiring that the magnetic field has a well defined power spectrum
also for $k\rightarrow 0$, we have
\begin{equation}
\int d^3{\bf r} \,|{\mathcal B}_{ii}({\bf r}) | < \infty.
\end{equation}
As we will show below, this inequality also ensures that
the average (mean) magnetic energy density
$\EEM=\langle{\bf B}({\bf x})\cdot{\bf B}({\bf x})\rangle/2$
is well defined in wavenumber space ${\bf k}$.

Let us now consider the spectral (Fourier) decomposition of our
stochastic magnetic field amplitudes\footnote{We define the Fourier
transform of the magnetic field, ${\bf B}({\bf x})$, with the following
normalization:
\begin{eqnarray}\label{fourier-modes}
{\bf B}({\bf k}) = \int d^3\mathbf x \,e^{i\mathbf k\cdot\mathbf x} {\bf B}(\mathbf x),\quad\quad
{\bf B}({\bf x}) = \frac{1}{(2\pi)^3}\int d^3\mathbf k \, e^{-i\mathbf k\cdot\mathbf x}{\bf B} (\mathbf k).
\end{eqnarray}}
${\bf B}({\bf k})$. Reality of $ {\bf B}({\bf x})$ implies
${\bf B}({\bf k}) = {\bf B}^\star (-{\bf k})$,
and due to statistical homogeneity the 2-point statistical average is of the form
\begin{eqnarray}
\langle B_i^\star({\bf k}) B_j({\bf k'}) \rangle = (2\pi)^3\delta^{(3)}(\mathbf k-\mathbf k') {\mathcal F}^{(B)}_{ij}({\bf k}),
\label{spectrum}
\end{eqnarray}
where $\langle\cdots\rangle$ again denotes ensemble average, but now
in wavenumber space. The matrix ${\mathcal F}^{(B)}_{ij}({\bf k})$ is called the three-dimensional (3D) power spectrum
of the magnetic field, and in fact, it is the Fourier transform of the
magnetic field two-point correlation function ${\mathcal B}_{ij}({\bf r})$ (see \App{FourierTwoPoint}):
\begin{eqnarray}
{\mathcal B}_{ij}(\mathbf r) &=& \frac{1}{(2\pi)^3}\int d^3\mathbf k \,e^{-i\mathbf k \cdot \mathbf r}{\mathcal F}^{(B)}_{ij}(\mathbf k),\label{B-ij}\\
{\mathcal F}^{(B)}_{ij}({\bf k}) &=& \int d^3\mathbf r \,e^{i\mathbf k \cdot \mathbf r} {\mathcal B}_{ij}(\mathbf r).
\label{spectral-density}
\end{eqnarray}
Translational invariance of the magnetic field is reflected in the
presence of the Dirac delta function $\delta^{(3)}({\bf k}-{\bf k}')$
on the right hand side of equation~(\ref{spectrum}).
For the general case, when the isotropic stochastic Gaussian magnetic
field has non-zero helicity ($M_{\rm H}(r) \neq 0$), the 3D spectrum
matrix components ${\mathcal F}^{(B)}_{ij}({\bf k})$ satisfy the following
reality conditions:
\begin{equation}
{\mathcal F}^{(B)}_{ij}({\bf k})={\mathcal F}^{(B)}_{ji}(-{\bf k}) = [{\mathcal F}^{(B)}_{ij}]^\star(-{\bf k})=[{\mathcal F}^{(B)}_{ji}]^\star({\bf k})\,.
\label{spectral-relations}
\end{equation}
Defining $P_{ij}({\bf {\hat k}}) = \delta_{ij} - {\hat k}_i {\hat k}_j$,
the projection operator onto the plane normal to $\bf k$ with
${\hat k}_i = k_i/k $, $k=|{\bf k}|$, the divergence-free condition
of the magnetic field
requests the following form for  ${\mathcal F}_{ij}({\bf k})$
\begin{eqnarray}
\frac{{\mathcal F}^{(B)}_{ij}({\bf k}) }{(2\pi)^3}= P_{ij}({\bf {\hat  k}})
\frac{\EM(k)}{4\pi k^2} + i \varepsilon_{ijl} {k_l}
\frac{\HM(k)}{8\pi k^2}. \label{4.1}
\end{eqnarray}
The most general form
for the function ${\mathcal F}_{ij}({\bf k})$ is
\begin{equation}
{\mathcal F}_{ij}({\bf k}) = P_{ij}({\bf {\hat k}}) {\mathcal F}_{\rm N}(k) + {\hat k}_i {\hat k}_j {\mathcal F}_{\rm L}(k) + i\varepsilon_{ijl} k_l {\mathcal F}_{\rm H}(k)\,.
\label{correlfourier}
 \end{equation}
The functions ${\mathcal F}_{\rm N}(k)$, ${\mathcal F}_{\rm L}(k)$, and
${\mathcal F}_{\rm H}(k)$ represent the normal, longitudinal, and
helical (antisymmetric) parts of the magnetic 3D spectrum, and
they are integrals over the corresponding function in real space
\footnote{\label{foot5}
The functions ${\mathcal F}_{\rm N}(k)$, ${\mathcal F}_{\rm L}(k)$, and
${\mathcal F}_{\rm H}(k)$ can be expressed in terms of $M_{\rm N}$, $M_{\rm L}$, and $M_{\rm H}$ as
\begin{eqnarray}\label{eFoots}
{\mathcal F}_{\rm L}(k) & = & 4\pi \int_0^\infty dr \, r^2 \Big[j_0(kr)M_{\rm L}+\frac{2j_1(kr)}{kr} (M_{\rm N}-M_{\rm L})\Big]\,,
\nonumber \\
{\mathcal F}_{\rm N}(k) & = & 4\pi  \int_0^\infty dr \, r^2 \Big[j_0(kr)M_{\rm N}+\frac{j_1(kr)}{kr}(M_{\rm L}-M_{\rm N})\Big]\,, \nonumber \\
{\mathcal F}_{\rm H}(k) &=& -8\pi  \int_0^\infty dr \, r^2 \Big [\frac{j_1(kr)}{kr}M_{\rm H}\Big]\,,
\end{eqnarray}
where $j_n(x)$ is the spherical Bessel function of order $n$.
The divergenceless condition of the magnetic field implies
$k_i{\mathcal F}_{ij}({\bf k}) = k_j {\mathcal F}_{ij}({\bf k})=0$,
and thus the longitudinal component ${\mathcal F}_{\rm L}(k)$
must be vanishing for the divergence-free field.} Using integrals given in Eqs. \ref{eFoots}, as well as the
spherical Bessel function identity $(x^2j_1(x))'=x^2j_0$, one sees
immediately that (\ref{div-condition}) implies ${\mathcal F}_{\rm L}(k) \equiv
0$ for a divergence-free field. The function $\EM(k)$ is determined by the
symmetric part of the magnetic field 3D spectrum, it is usually referred
to as the spectral energy density, $\int dk \, \EM(k) = \EEM$,
see \Sec{RMSenergy}, below, and it can be expressed in terms of the Fourier
transform of the normal component of the correlation function given
in equation~(\ref{correlreal}), ${\mathcal F}_{\rm N}(k)$ as $\EM (k)  = {{\mathcal F}_{\rm N}(k)}/({4\pi k^2})$.
The function $\HM(k)$ is the antisymmetric part of the 3D magnetic field
spectrum, usually referred to as the spectral density of the magnetic helicity,
$\int dk \HM(k) = \HHM$, see \Sec{CurrentHelicity} below.
It can be expressed in terms of the helical component of the Fourier
transform of the correlation function equation~(\ref{correlreal}),
$\HM(k)={{\mathcal F}_{\rm H}(k)}/({2\pi k})$.
The magnetic field spectral energy and helicity densities are typically
given by simple power laws in a certain wavenumber range; generally
different spectral ranges are characterized by different spectral indices,
\begin{equation}
\EM(k) \propto k^{n_E} , \quad \HM(k) \propto k^{n_{\rm H}}.
\label{powerlaw}
\end{equation}
Of particular interest are the spectral shapes at large length scales,
i.e., small wavenumbers, and, from now on, whenever unspecified,
``spectral index'' refers to the spectral index at the large-scale
asymptotics.
These spectral indices $n_E$ and $n_{\rm H}$ determine the shapes of spectral
energy and helicity of the magnetic field
{\it only} at large length scales (small wavenumbers).
They are defined by $\lim_{k \rightarrow 0}\EM/k^{n_E}=\mbox{finite}$
and $\lim_{k\rightarrow 0}\HM/k^{n_{\rm H}}=\mbox{finite}$.

\subsection{Mean and rms energy densities}
\label{RMSenergy}

The mean magnetic energy density per unit volume, $\EEM$, is given by
\begin{eqnarray}
\EEM &=& \langle \rho_{\rm M}({\bf x}) \rangle =
\frac{1}{2} \langle |{\bf B}({\bf x})|^2 \rangle =
\frac{1}{2} {\mathcal B}_{ii}(0) = M_{\rm N}(0) + \frac{1}{2} M_{\rm L}(0) \,.
\label{energy}
\end{eqnarray}
Since $M_{\rm N}(0)=M_{\rm L}(0)$,\footnote{The equality can be shown as
follows. We rotate the coordinate frame ${\bf x} \rightarrow {\bf x}'$,
so that the axis ${\bf e}_1$ is now in the direction of ${\bf r}$.
Then there will only be two independent components of the matrix
${\mathcal B}'_{ij}({\bf r}')$.
These components correspond to the normal and longitudinal components,
and they must be equal to each other at ${\bf r}=0$.
This is also obtained from \eqref{div-condition}, assuming $M_L'(r)|_{r=0}<\infty$.
See \cite{monin-yaglom} for details.} we have
\begin{equation}
M_{\rm N}(0) = M_{\rm L}(0) = \frac{2}{3} \EEM.
\end{equation}
Note again that the mean energy is independent of the helicity given by
$M_{\rm H}(r)$ or ${\mathcal F}_{\rm H}(k)$.
The quantity ${\mathcal B}_{ii}(0)$ is given by the trace of the
3D spectrum, ${\mathcal F}_{ii}({\bf k})$ (which is continuous at
$\mathbf k\to 0$, provided the magnetic energy density does not
diverge at infinity), and thus ${\mathcal B}_{ii}(0) =
\delta_{ij} \lim_{{\bf r}\to 0} {\mathcal B}_{ij}({\bf r})$.
We have
\begin{eqnarray}\label{avgEne}
 \EEM= \frac{1}{2(2\pi)^3} \delta_{ij} \lim_{{\bf r} \rightarrow 0} \int d^3
{\bf k} \, e^{-i{\bf k}\cdot {\bf r}} {\mathcal F}^{(B)}_{ij}({\bf k})
= \int dk\, \EM(k)\,.
\end{eqnarray}
Thus $\EM(k)$
describes the distribution of the magnetic energy density in wavenumber
space, which justifies its definition as the \textit{spectral energy
density of the magnetic field}.\footnote{
The symmetric spectra $M_{\rm N}(r)$ and  $M_{\rm L}(r)$ in terms of $\EM(k)$ and
the spherical Bessel functions $j_n(x)$ can be written as:
\begin{eqnarray}
M_{\rm L}(r) = 2 \int_0^\infty dk \EM(k)\frac{j_1(kr)}{kr},\quad\quad
M_{\rm N}(r) =   \int_0^\infty dk \Big[j_0(kr)-\frac{j_1(kr)}{kr} \Big ] \EM(k).
\end{eqnarray}
These are just the inverse Fourier transforms of the expressions
in footnote~\ref{foot5} for the case ${\cal F}_{\rm L}=0$.}
The requirement that the magnetic energy density converges toward infinity
(${k} \rightarrow 0$) implies that $n_E > -1$.
A spectrum with $n_E \rightarrow -1$, i.e., $E(k) \propto k^{-1}$ is
a {\it scale-invariant} spectrum.
Such a magnetic field can be generated during the inflationary epoch;
for a review see \cite{Subramanian:2015lua}
and references therein.

We denote the integral $\int_0^\infty dk \, \EM(k)$ by $\int dk \, \EM(k)$.
In reality, the power law (with different spectral shapes in the different
regimes) for $\EM(k)$ only holds below the magnetic field cutoff scale
$k_D$, i.e.\ $\EM(k) \rightarrow 0$  for $k>k_D$.\footnote{More precisely
for $k>k_D$ the magnetic field spectral energy density experiences the
exponential cutoff, $\EM(k) \propto e^{-(k/k_D)^2}$ for $k>k_D$.}

The magnetic root mean square (rms) energy density is defined as
\begin{equation}
\EEM^{\rm rms} = \big[\langle |\rho_{\rm M}({\bf x})|^2 \rangle\big]^{1/2}
= \frac{1}{2}\sqrt{\langle |\mathbf B(\mathbf x)|^4\rangle}.
\label{rho}
\end{equation}
Hence,
\begin{equation}
\EEM^{\rm rms} = \frac{1}{2}\lim_{\mathbf r\to 0} \sqrt{{\mathcal B}_{ii,jj}(\mathbf r)},
\label{rho4point}
\end{equation}
where ${\mathcal B}_{ij,lm}$ is the  four-point correlation function of the magnetic field, defined as
\begin{equation}\label{e:4pt-2pt}
{\mathcal B}_{ij,lm}(\mathbf r) = \langle B_i(\mathbf x)B_j(\mathbf x)B_l(\mathbf x+\mathbf r)B_m(\mathbf x+\mathbf r)\rangle.
\end{equation}
As we consider a Gaussian magnetic field, we can apply Wick's theorem
and express the four-point correlation function in terms of the two-point
correlation functions,
\begin{equation}
{\mathcal B}_{ij,lm}({\bf r})={\mathcal B}_{ij}(0){\mathcal B}_{lm}(0) + {\mathcal B}_{il}({\bf r}) {\mathcal B}_{jm}({\bf r})+{\mathcal B}_{im}({\bf r}) {\mathcal B}_{jl}({\bf r}).
\label{wick}
\end{equation}
We emphasize that the first term on the rhs of equation~(\ref{wick}) is usually
discarded (see, for example, ref.~\cite{Ballardini:2014jta}) because it
cannot be obtained through ${\bf k}$-space considerations.

From the definition of the 4-point function it is clear that,
\begin{equation}\label{sym}
{\mathcal B}_{ij,lm} ({\bf r}) = {\mathcal B}_{lm, ij} ({\bf r}) = {\mathcal B }_{ji,lm}({\bf r})={\mathcal B}_{ij,ml}({\bf r}).
\end{equation}
We now calculate the trace of (\ref{e:4pt-2pt}) and obtain,
\begin{equation}
\mathcal{B}_{ii,ll}(\mathbf{r})=9M_{\rm N}^2(0)+4M_{\rm N}^2(r)+2M_{\rm L}^2(r)+4r^2M_{\rm H}^2(r)\,.
\label{rms-energy}
\end{equation}
Reconstructing equation~(\ref{rms-energy}) from the magnetic field
rms energy density power spectrum must be done with caution: the first
constant term $9M_{\rm N}^2(0)$ will be missing when naively taking the direct
Fourier transform of ${\mathcal B}_{ii,ll}(k)$.
To find the rms magnetic energy density, we now take the limit of
${\mathcal B}_{ii,ll}(\mathbf r)$ as ${\bf r}\to 0$. Equation~(\ref{rho4point}) then gives (see \App{a:B})
\begin{equation}
\EEM^{\rm rms} = \frac 12 \sqrt{15} M_{\rm N}(0) = \sqrt{\frac{5}{3}} \EEM.
\label{EEMrms}
\end{equation}
We conclude that {\it the rms magnetic energy density
$\EEM^{\rm rms}$, just like the average magnetic energy
density $\bar\rho_{\rm M} =\EEM$, does not depend on magnetic
helicity}.\footnote{Note that ref.~\cite{Ballardini:2014jta} gives a potentially
misleading expression in their equation~(3.4), implying the presence
of a helical contribution to the rms energy density.
That expression is only the $\mathcal{O}(k^0)$ term of the
fuller expressions given in their Appendix~B, and could be misunderstood.
The bottom left panel of their figure~1 indicates
positive and negative contributions from small and large $k$,
suggesting vanishing rms energy density, although it is not explicitly mentioned.}
For this it is important that we define the helical component as
$\epsilon_{ijl}r^lM_H(r)$ so that it vanishes as ${\bf r}\to 0$,
which it must as a consequence of its antisymmetry under parity.

\subsection{Characteristic length scales}

In this section we define the relevant characteristic length scales
for the magnetic fields that are related to the correlation and damping
length scales.
We distinguish between statistically relevant length scales and
smoothing length scales: Statistical length scales are determined
fully by the magnetic field configuration and the  properties of the
plasma (viscosity, diffusivity, etc), while smoothing length scales
are introduced for convenient interpretation and {\it normalization}
purposes. In cosmology, it is often convenient to use a smoothing length of 1~Mpc
since an amplitude of about $10^{-9}$~Gauss on this scale is required
to form the observed magnetic fields in clusters by pure contraction,
without a dynamo mechanism \cite{Dolag:1999,Dolag:2002}. Depending on the magnetic field generation mechanism, this scale of
1~Mpc can be substantially larger than the magnetic correlation length
(e.g., for magnetic fields generated during the electroweak phase transition), while for other mechanisms, 1~Mpc can be a small fraction of the magnetic
correlation length (e.g., for inflationary magnetic fields).
We discuss the normalization aspects in section~\ref{sec:normalization}.

\begin{figure}
\begin{center}
\includegraphics[width=.7\columnwidth]{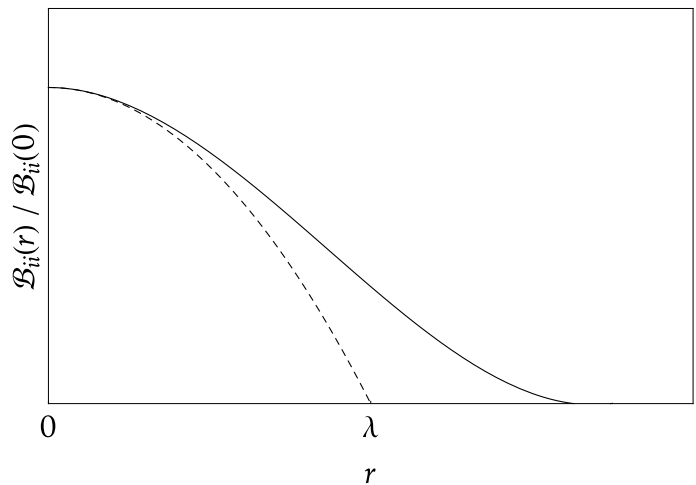}
\end{center}
\caption{
The damping scale is the length segment that is cut off on the abscissa
by a parabola that can be locally fit to the $B_{ii}(r)$ curve at its apex
($r = 0$), see also ref. \cite{monin-yaglom}.
}\label{damping}
\end{figure}

Let us first consider the characteristic length scales that are constructed
solely from integration of the spectral magnetic energy. The magnetic integral scale is defined as \cite{monin-yaglom}
\begin{equation}\label{Int-LenSc}
\xiM(t)=\kM^{-1}=\frac{\int_0^\infty dk\,k^{-1}
E_{\rm M}(k,t)}{{\mathcal E}_{\rm M}(t)}.
\end{equation}

There are various possibilities to construct additional
length scales from the second derivative of the magnetic two-point correlation function.
In particular, we define the {\it differential} scale for the magnetic
field as \cite{monin-yaglom}
\begin{equation}
\lambda_{\rm M} =\left| \frac{{\mathcal B}_{ii}(0)}{{\mathcal B}_{ii}^{\prime\prime}(0)}\right|^{1/2}.
\label{lambdam}
\end{equation}
The differential length scale characterizes the scale beyond which
correlations between two points are significantly washed out.
In \Fig{damping} we illustrate the meaning of the differential length scale.
The wavenumber corresponding to the differential length scale
$\lambda_{\rm M}$ is known as the magnetic {\it Taylor} microscale wavenumber,
\begin{equation}
k^2_{\rm MT}=\frac{\int_0^\infty dk\,k^2 \EM(k)}\EEM
\label{kMT}
\end{equation}
with $\lambda_{\rm M}=k^{-1}_{\rm MT}$. In Kolmogorov's hydrodynamic turbulence theory, dissipation is described
through the mean kinetic energy dissipation per unit mass: $\epsK=2\nu\int dk \, k^2 \tEK(k)$, where $\nu$ is the kinematic viscosity, and the tilde on $E$ (and later also on ${\cal E}$)
is used here and below to indicate energies or spectral energies per unit
mass, applying normalization by $p+\rho$, where $p$ and $\rho$ denote the pressure and the energy density of the plasma.
Thus, analogously to $\epsK$, the magnetic energy dissipation is given by
$\epsM \propto  2\eta\int dk \, k^2 \tEM(k)$.
The magnetic dissipation wavenumber is defined through
\begin{equation}
\kMD^4 = \frac{\epsM}{\eta^3} = \frac{2\int_0^\infty dk\,k^2 \tEM(k)}{\eta^2},
\label{kM}
\end{equation}
where $\eta$ is the magnetic diffusivity.
In the present case, however, the turbulent flow is driven entirely
by the magnetic field and the magnetic energy spectrum follows a
weak turbulence (WT) spectrum of the form
\begin{equation}
\tEM(k)=C_{\rm WT} \, (\epsM\vA\kM)^{1/2}\,k^{-2},
\label{EWT}
\end{equation}
where $C_{\rm WT}\approx1.9$ is a Kolmogorov-type constant for
magnetically dominated turbulence \cite{Brandenburg:2014mwa}, and
$\vA=(2\tEEM)^{1/2}$ is the Alfv\'en speed.
Defining now a $k$-dependent Lundquist number as
$\Lu(k)=\vA(k)/\eta k$ with $\vA^2(k)=2k\EM(k)$ and defining
a WT dissipation wavenumber through $\Lu(\kWT)=1$, we find\footnote{
Inserting $k=\kWT$ into $\Lu(\kWT)=1$, we have
$1=(\eta\kWT)^{-1}\sqrt{2\kWT C_{\rm WT} \, (\epsM\vA\kM)^{1/2}\,\kWT^{-2}}$.
Raising this to the fourth power and solving for $\kWT$ yields
$\kWT^6=(2\,C_{\rm WT})^2 \epsM\vA\kM/\eta^4$.
The combination $\epsM/\eta^3$ is just $\kMD^4$ from equation \Eq{kM}
and $\Lu\equiv\vA/\eta\kM$ is now defined as a $k$-independent
Lundquist number.}
\begin{equation}
\kWT^6=(2C_{\rm WT})^2\,\Lu\,\kM^2\,\kMD^4.
\label{kWT}
\end{equation}
In section~\ref{sec:Cosmology} below, we show the values of some of the
characteristic length scales during the MHD decay.

\subsection{Current and magnetic helicity}
\label{CurrentHelicity}

The antisymmetric part of the magnetic field spectrum, $M_{\rm H}(r)$, is
related to the magnetic helicity.
It is also related to the current helicity, but this is usually less crucial
than the magnetic helicity, which satisfies a conservation equation.
The mean magnetic helicity density over a volume $V$ is defined as
\begin{eqnarray}
\HHM &=& \lim_{V\rightarrow \infty} \frac 1V\int_V d^3 {\bf x}\; \mathbf A\cdot\mathbf B
=\lim_{V\rightarrow \infty}\frac 1V\int_V d^3 {\bf x}\, \Big(\mathrm{curl}^{-1}\mathbf B\Big)\cdot\mathbf B\,.
\label{magnetic-helicity}
\end{eqnarray}
In the mathematical literature, this is called a \textit{generalized asymptotic form of the Hopf invariant} \cite{AK92},
or the \textit{measure of line linkage} of the $\mathbf{B}$ field.
This quantity cannot be defined locally, and in a realistic situation
the infinite volume
should be understood as
a 3D volume where the magnetic field
is determined.

Note that the form of $\HHM$ is gauge-dependent, unless
the domain is periodic \cite{gauge} or the normal component of $\mathbf B$
vanishes at infinity \cite{Kunze:2011bp}.
However, this only affects the magnetic helicity spectrum at the
largest scales \cite{SB06}.
Let us also note that in hydrodynamics, the kinetic helicity of the
velocity field $\mathbf v$ is
\begin{equation}
{\mathcal H}_K = \frac 1V\int_V d^3\mathbf x \,\Big[\mathbf v\cdot \big(\mathbf \nabla \times \mathbf v\big)\Big].
\end{equation}
By analogy, it is customary to define the current helicity of the
magnetic field $\mathbf B$ as
\begin{equation}
{\mathcal H}_C = \frac 1V\int_V d^3\mathbf x \, \Big[\mathbf B\cdot \big (\mathbf \nabla \times\mathbf B\big)\Big],
\end{equation}
which is gauge-invariant.

Replacing $\frac{1}{V} \int d^3 {\bf x} $ by an ensemble average, $\langle\cdots\rangle$,
and using the definition of $\HM(k)$ in equation~(\ref{4.1}), we find that the magnetic
helicity can be written as
\begin{equation}\label{maghel}
\HHM = \int dk\, \HM(k)\,,
\end{equation}
while the current helicity can be expressed as
\begin{equation}\label{curhel}
{\mathcal H}_C = \int dk\,k^2 \HM(k)\,.
\end{equation}

\subsection{Smoothed magnetic field and helicity}
\label{sec:normalization}

We also define the smoothed magnetic field amplitude and magnetic helicity density
over a smoothing length scale $\sim\lambda$ using a Gaussian window function $e^{-\lambda^2k^2}$,
\begin{equation}\label{smoothB}
B_\lambda^2=\int\EM(k)\,e^{-\lambda^2k^2}\,dk,\quad H_\lambda=\int\HM(k)\,e^{-\lambda^2k^2}\,dk,
\end{equation}
which can be related to the average energy by eliminating the normalization. This gives rise to
\begin{equation}\label{eNorm1}
\frac {B_\lambda^2} {\EEM} =\frac {\ds{\int\EM(k)\,e^{-\lambda^2k^2}\,dk}} {\ds{\int\EM(k)\,dk}} \, ,\quad
\frac {H_\lambda}   {\HHM} =\frac {\ds{\int\HM(k)\,e^{-\lambda^2k^2}\,dk}} {\ds{\int\HM(k)\,dk}} \,.
\end{equation}

\section{The realizability condition}
\label{sec:HelRealCond}

If the magnetic diffusivity is very small, as it is for the highly
conductive plasma of the early universe, it can be shown that the
\textit{magnetic helicity is conserved}.
This leads to the theorem \cite{Arnold2014}
\begin{eqnarray}
&&\mbox{\it The eigenfield of $\mathrm{curl}^{-1}$ corresponding to the
eigenvalue $L$ of the largest modulus} \nonumber \\
&&\mbox{\it has minimum energy in the class of divergence free fields
obtained from}\\
&&\mbox{\it the eigenfield under the action of volume-preserving diffeomorphisms.}\nonumber
\end{eqnarray}
In other words, if $L_-$ and $L_+$ denote the smallest and largest
eigenvalues of the $\mathrm{curl}^{-1}$ operator respectively, with
$L_-<0< L_+$, then for every divergence-free field $\textbf{B}$, we have
\begin{equation}
L_-|\mathbf{B}(\mathbf{x})|^2\leq\left(\mathrm{curl}^{-1}\mathbf{B}\right)
\cdot\mathbf{B}\leq L_+|\mathbf{B}(\mathbf{x})|^2,
\end{equation}
which implies
\begin{equation}
\label{e39b}
\left|\frac{\left(\mathrm{curl}^{-1}\mathbf{B}\right)
\cdot\mathbf{B}}{L}\right|\leq|\mathbf{B}(\mathbf{x})|^2.
\end{equation}
Here $L=\max\{-L_-,L_+\}$ is the larger  of the moduli of the two eigenvalues.
Taking an ensemble average leads to
\begin{equation}\label{e40}
\left<|\mathbf{B}(\mathbf{x})|^2\right>=
2\EEM\geq\left|\frac{\left<\left(\mathrm{curl}^{-1}\mathbf{B}\right)
\cdot\mathbf{B}\right>}{L}\right|=\left|\frac{\HHM}{L}\right|.
\end{equation}
For a field with eigenvalue $\pm L$, the equality sign holds but in general we
 have $|\HHM|\leq2|L|{\mathcal E}_M$.

It is justified to assume that $|L|$ is of the order of the
magnetic integral length scale, $|L|\lsim \xiM$, since this is the
maximal length scale that can be associated with the size of a domain.
This leads to the well-known realizability condition
\begin{equation}
|\HHM|\leq2\xi_M\EEM.
\end{equation}
This equation, together with the definitions in equations~\eqref{avgEne},
\eqref{Int-LenSc} and \eqref{eNorm1}, leads to
\begin{equation}\label{ereal1}
\left|H_ \lambda\right| \leq f(\lambda)B_\lambda^2,
\end{equation}
where
\begin{equation}\label{ereal2}
f(\lambda)=2 \left| \frac
{\ds{{\textstyle \int\HM(k)\,e^{-\lambda^2k^2}\,dk}}}
{\ds{{\textstyle \int\HM(k)\,dk}}} \right|
\left/ \frac
{\ds{\textstyle \int\EM(k)\,e^{-\lambda^2k^2}\,dk}}
{\ds{\textstyle \int k^{-1}\EM(k) \,dk}} \,.\right.
\end{equation}
In the examples discussed below, which are all for the fully helical
case with $k\HM=2\EM$, we show that for $\lambda\ll\xiM$, we have $f(\lambda) \simeq \xiM$,
while for $\lambda\gg\xiM$, we have $f(\lambda)\simeq\lambda$.
This is simply because in the Batchelor subinertial range, we have
$\HM(k)\propto k^3$, while $\EM(k)\propto k^4$ (see section~\ref{3.2} below), so
$H_\lambda\propto\lambda^{-4}$ and
$B_\lambda^2\equiv E_\lambda\propto\lambda^{-5}$.

\subsection{Normalized magnetic field and helicity}
\label{sec:DiffShapes}

To satisfy the condition \eqref{ereal1} on all scales, we need to ensure that the realizability condition embodied by
equation~\eqref{ereal2} holds on all scales.
For this purpose we analyze magnetic fields with different spectral energy
distributions employing statistical properties of random fields from the
theory of turbulence.
In this approach, we split the stochastic field into its large-scale
and small-scale spectra in wavenumber space.
We define the large-scale spectrum at wavenumbers smaller than the
integral scale of the random field $k<k_1=\xiM^{-1}$. By comparison, the small-scale spectrum, corresponding to the inertial range of turbulence, occurs at $k>k_1$.
We proceed with detailed calculations for two relevant cases.

\subsection{Batchelor spectrum}\label{3.2}

\begin{figure}
\begin{center}
\includegraphics[width=.7\columnwidth]{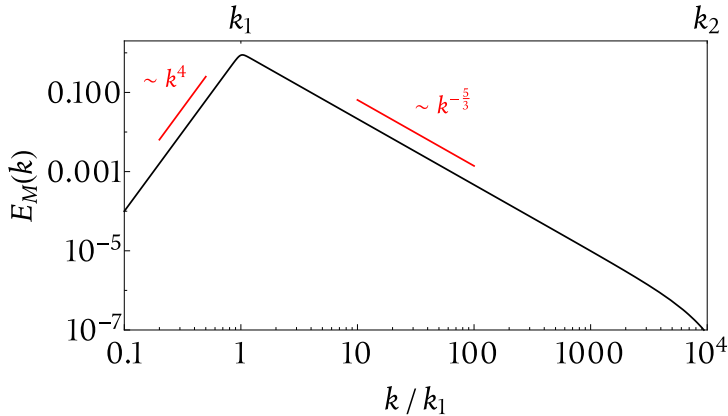}
\end{center}
\caption{\label{f:BatSpec} The spectral distribution of the energy of
random magnetic fields matching the Batchelor spectrum at large scales
($k<k_1$) and Kolmogorov spectrum at small scales ($k>k_1$) in arbitrary units.
The dissipative cutoff occurs at $k>k_2$.
}\end{figure}

\begin{figure}
\begin{center}
\includegraphics[width=.7\columnwidth]{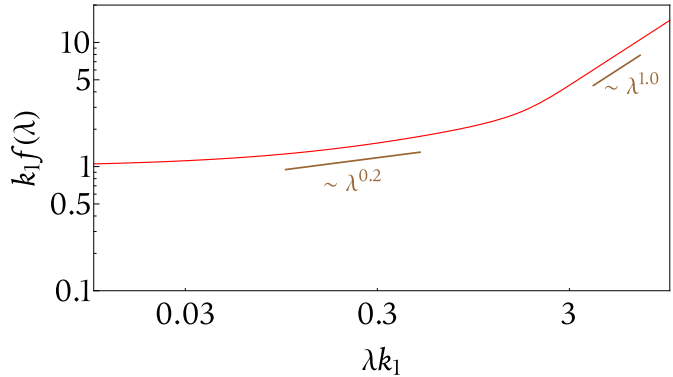}
\end{center}
\caption{The function $f(\lambda)$ defined in
\eqref{ereal2}, for a Batchelor spectrum at large scales.
We assume a maximally helical field where the inequality (\ref{e43})
is saturated.
For $\lambda k_1 \ll 1$, we have $f(\lambda)\to k_1^{-1}$, while for
$\lambda k_1 > 1$, $f(\lambda)\sim\lambda$.
For intermediate values, $\lambda k_1 \approx 0.3$, we find
$f(\lambda)\sim \lambda^{0.2}$.
}\label{FigLamBata}
\end{figure}

The large-scale part of the spectrum of the turbulent fluctuations is often described by a Batchelor spectrum.
In this case the energy spectrum at large scales grows as $\sim k^4$. This behavior at large scales is a consequence of
causality \cite{Durrer:2003ja}: If the correlation function in real
space has finite support, and since correlations have been generated in the
finite past and can spread out no faster than with the speed of light,
its Fourier transform, ${\cal F}_{ij}({\bf k})$, must be analytic.
The nonanalytic prefactor $P_{ij}({\bf {\hat k}})$ then requires that
${\cal F}_{\rm N} \propto k^2$ and hence $\EM\propto k^4$.
On scales where the spectrum has already been affected by turbulence,
i.e., above the integral wavenumber (in the following referred to as $k_1$),
the spectral energy decreases according to the classical Kolmogorov exponent
$k^{-5/3}$, often described as the inertial range of turbulence.
At length scales smaller than some dissipative
length scale (with corresponding wavenumber defined as $k_2$)
the spectral energy undergoes an exponential cutoff.
The spectral distribution of the energy of random magnetic field can be modeled as
\begin{equation}\label{eBatSpec}
\EM(k)\propto \frac{k^4\,\exp[-(k/k_2)^2]}
{\left[1+(k/k_1)^{(5/3+4)q}\right]^{1/q}} ~,
\end{equation}
and $\HM(k)=2k^{-1}\EM(k)$. We use $q=5$ to make the transition between
the two subranges sufficiently sharp, see \Fig{f:BatSpec}.
Note that the correlation length $\xiM$ for such a spectral shape is
\textit{finite}. The function $f(\lambda)$ is plotted in \Fig{FigLamBata}.

Expanding the field in terms of the polarization basis, on can write the
magnetic energy spectrum as the sum of two positive definite contributions
and the magnetic helicity spectrum is then related to the difference of
these two contributions \cite{Durrer:2003ja,Brandenburg:2004jv}.
This leads directly to the realizability condition in wavenumber space,
\begin{equation}\label{e43}
\left|\HM(k)\right| \leq2k^{-1}\EM(k) \,.
\end{equation}
Note that a maximally helical magnetic field for which this inequality
becomes an equality, is either in a purely positive or in a purely
negative helicity configuration.

There is a special case when the integral scale of the random field
matches the spectral cutoff wavenumber $k_1=k_2$.
In this case, the initial increase of the spectral energy at large scales
is followed by an exponential cutoff at the integral scale, where the
energy reaches a maximum.
Hence, no turbulence occurs past the integral scale and a laminar regime
dominates.
Such a situation can be realized at very low Reynolds number, which is not
relevant to cosmology.
For cosmological applications the inertial range of the spectrum, where
spectral energy decays with wavenumber in $k_1 < k < k_2$ is an important
component contributing to the general form of the random magnetic field
configuration.

\subsection{Scale-invariant spectrum}

\begin{figure}
\begin{center}
\includegraphics[width=.7\columnwidth]{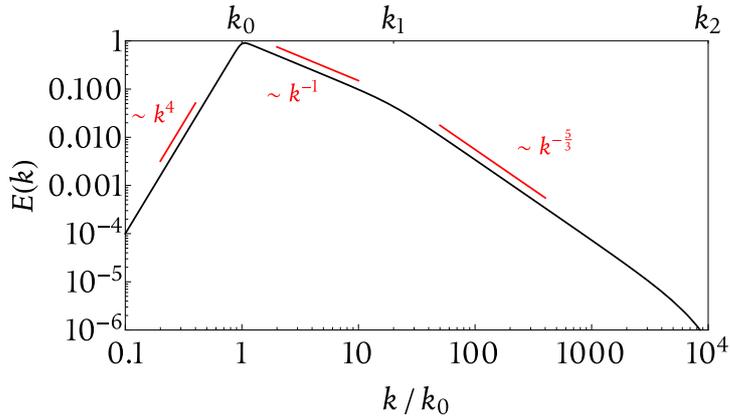}
\end{center}
\caption{The scale-invariant spectrum, with a $k^4$ dependence at low $k$.
We have chosen $k_1=20k_0$.
}\label{ScInvSpec}
\end{figure}

A scale-invariant spectrum has $E(k)\sim k^{n_E}$, with
$n_E\rightarrow-1$ for $k_0<k<k_1$.
For length scales smaller than $\sim k_1^{-1}$, turbulence is fully developed and the spectrum may be a Kolmogorov
spectrum proportional to $k^{-5/3}$ or a WT spectrum proportional
to $k^{-2}$, as given by \Eq{EWT}.
However if $n_E=-1$ at very large scales, the correlation length is unbounded
and the integral proportional to $\int dk\,k^{-1}E(k)$ does not converge.
We therefore cannot use equation~\eqref{e43} directly. To deal with this situation, we have to modify the spectrum at very
large scales for $k<k_0$.
In the inflationary case we may consider this to be the horizon scale at
the beginning of inflation \cite{Kahniashvili:2016bkp}. We assume the spectrum to have a $k^4$ dependence at
$k<k_0$, a $k^{-1}$ intermediate range for $k_0<k<k_1$,
a $k^{-5/3}$ inertial range for $k_1<k<k_2$, and an exponential cut-off
for $k>k_2$.
We can model it as (see  \Fig{ScInvSpec})
\begin{equation}\label{eScInvSpec}
\EM(k) \propto \frac{k^{-1}\,\exp[-(k/k_2)^2]}
{\left[1+(k/k_0)^{-(4+1)q}+(k/k_1)^{-(\tilde{n}_{\rm E}+1)q}\right]^{1/q}},
\end{equation}
where we choose again, $q=5$. We consider the fully helical case,
$\HM(k)=2k^{-1}\EM(k)$,
and $\tilde{n}_{\rm E}$ is chosen to be either $-5/3$ or $-2$.
Owing to the Batchelor subinertial range for $k<k_0$,
the correlation length is always finite, and we can use equation~\eqref{e43}
for $k\to0$. The function $f(\lambda)$ is plotted in \Fig{fLamScInv} for different
values of $k_1/k_0$ between $1$ and $10^4$, comparing two values
for the spectral inertial range exponent $\tilde{n}_{\rm E}$ of
$-2$ and $-5/3$.
The difference between these cases with different $\tilde{n}_{\rm E}$
is significant only for $k_0\lambda\ll1$ and if $k_1/k_0$ small.
For large values of $k_1/k_0$, there is now a clear $\lambda^{1/3}$
subrange for $0.1<k_0\lambda<1$.
As shown in the inset of \Fig{fLamScInv}, this slope emerges non-trivially
from both $E_\lambda$ and $H_\lambda$ being non-power laws.

\begin{figure}\begin{center}
\includegraphics[width=.47\columnwidth]{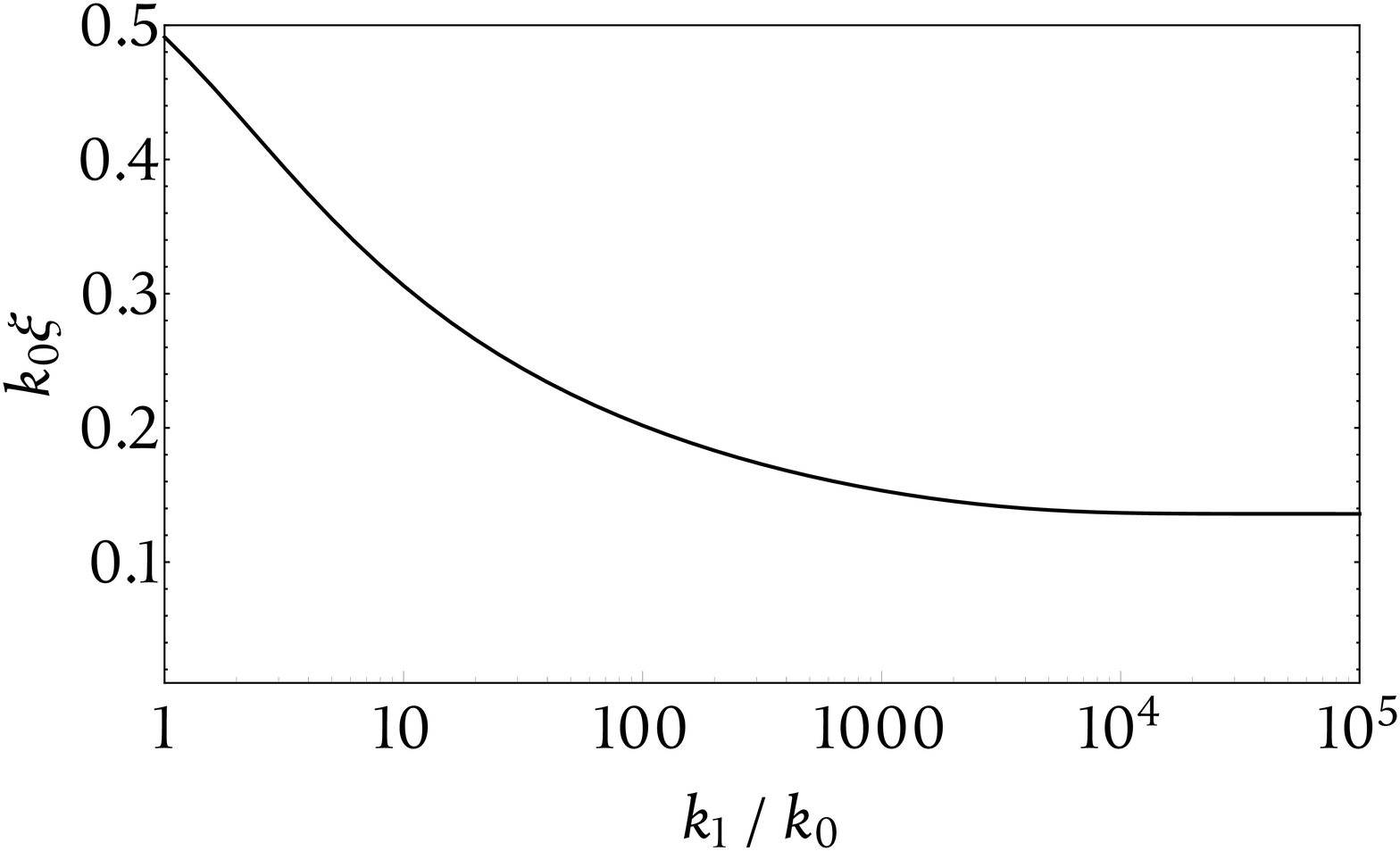}
\includegraphics[width=.47\columnwidth]{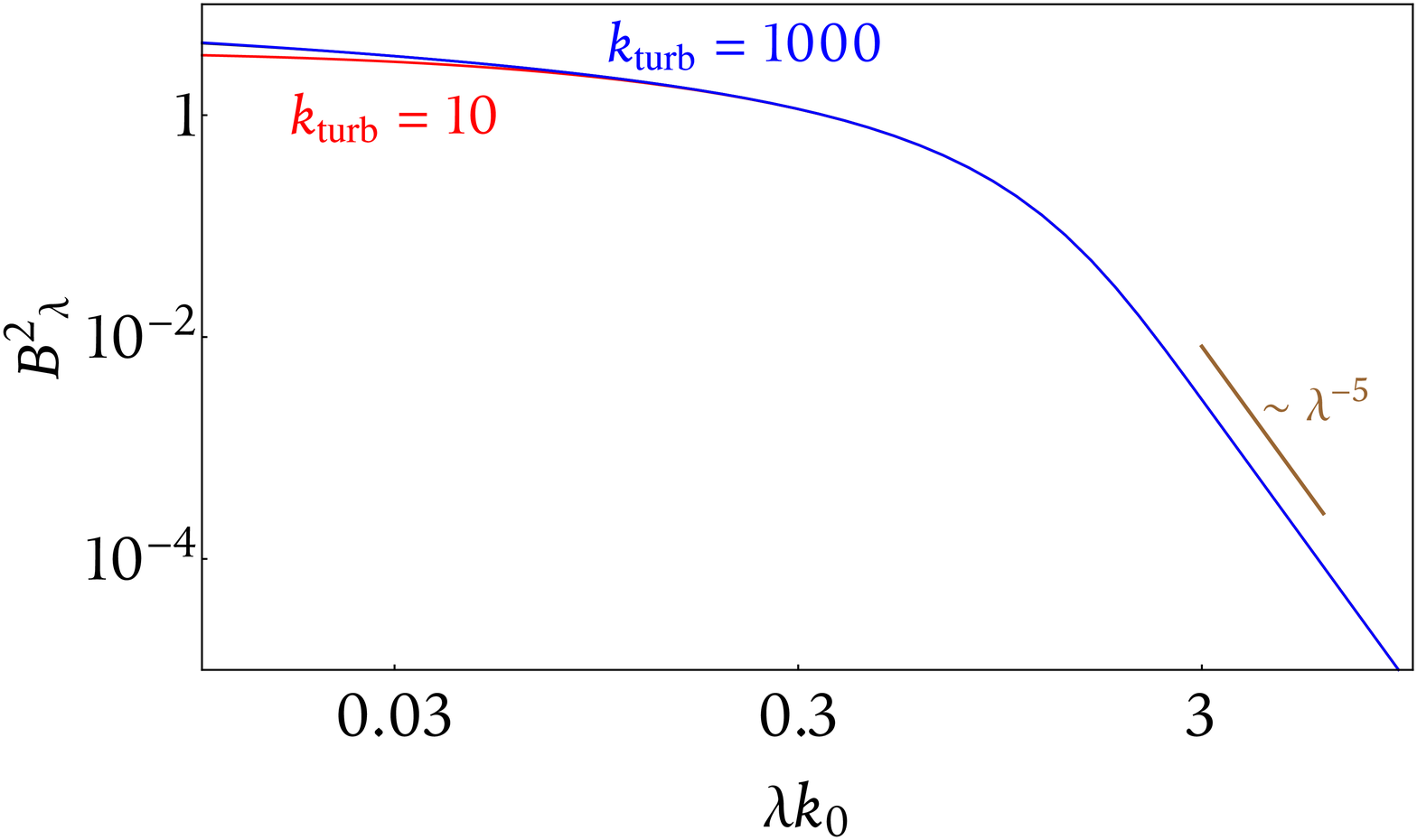}
\end{center}\caption[]{
Left: We show $k_0\xiM(k_1)$, as a function of $k_1/k_0$.
Right: $B_\lambda^2$ for $k_{\rm turb}/k_0\equiv k_1/k_0=10$ (red)
 and 1000 (blue).
}\label{ppspec_all}\end{figure}

\begin{figure}
\begin{center}
\includegraphics[width=.8\columnwidth]{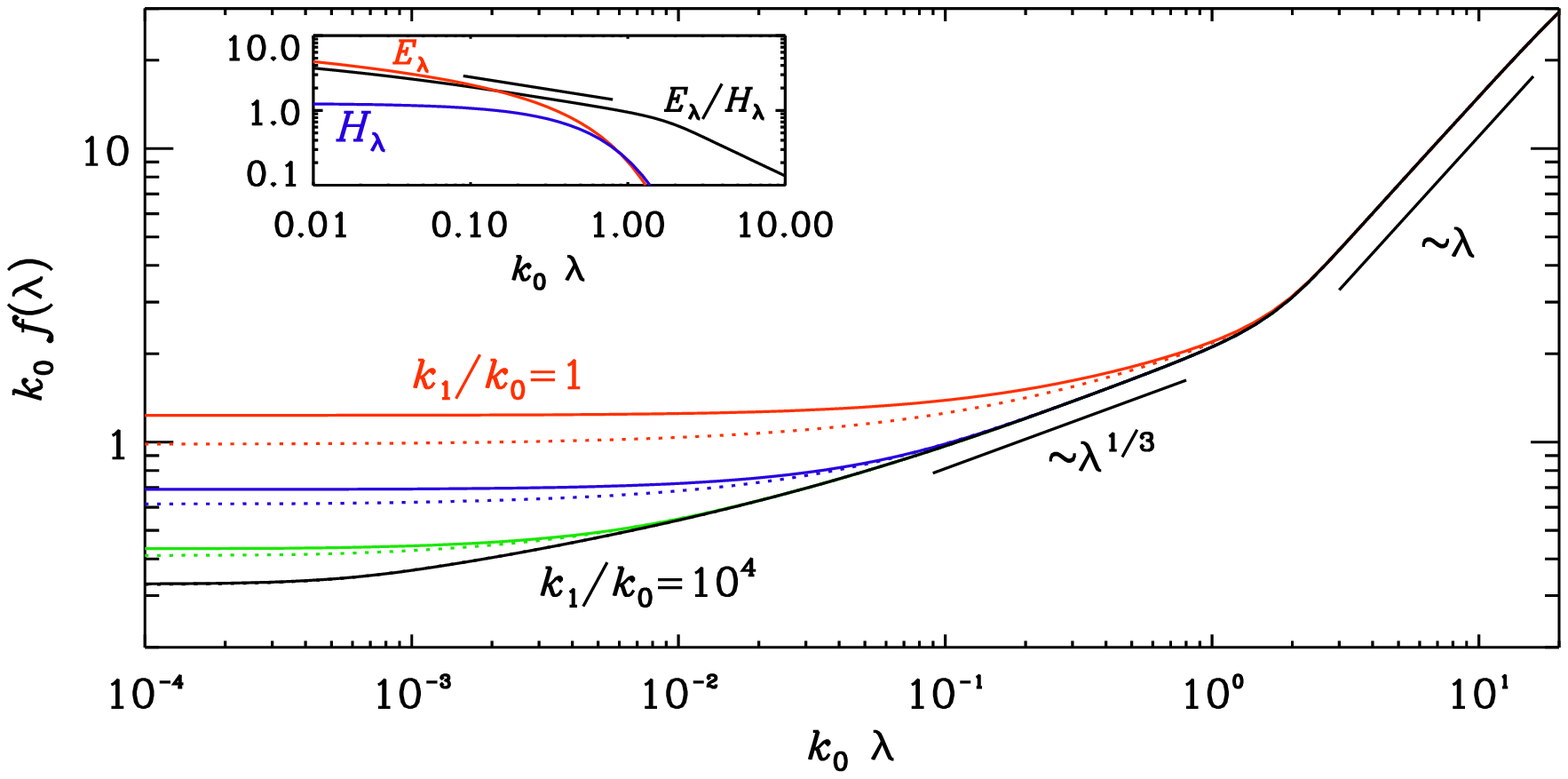}
\end{center}
\caption{The function $f(\lambda)$ for a scale-invariant spectral subrange
between $k_0$ and $k_1$, followed by a turbulent inertial range with
either $\tilde{n}_{\rm E}=-2$ (solid lines) or $\tilde{n}_{\rm E}=-5/3$
(dotted lines) for $k_1/k_0=1$ (red), $10$ (blue), $10^2$ (green),
and $10^4$ (black).
The inset shows that the $1/3$ slope emerges non-trivially from both
$E_\lambda$ and $H_\lambda$ being non-power laws.
}\label{fLamScInv}
\end{figure}

Inflation-generated magnetic fields have a large integral scale,
because it becomes exponentially amplified by a factor of $\sim e^{60}$. Turbulence develops and gradually leads to a $k^{-5/3}$ or $k^{-2}$ spectrum, followed by an exponential cutoff at a damping wavenumber $\kMD$ or $\kWT$.

The dependence of the smoothed magnetic field on the smoothing scale $\lambda$ is shown in the right panel of \Fig{ppspec_all}.

\section{Numerical simulations}
\label{sec:Numerical}

\begin{figure}
\begin{center}
\includegraphics[width=.8\columnwidth]{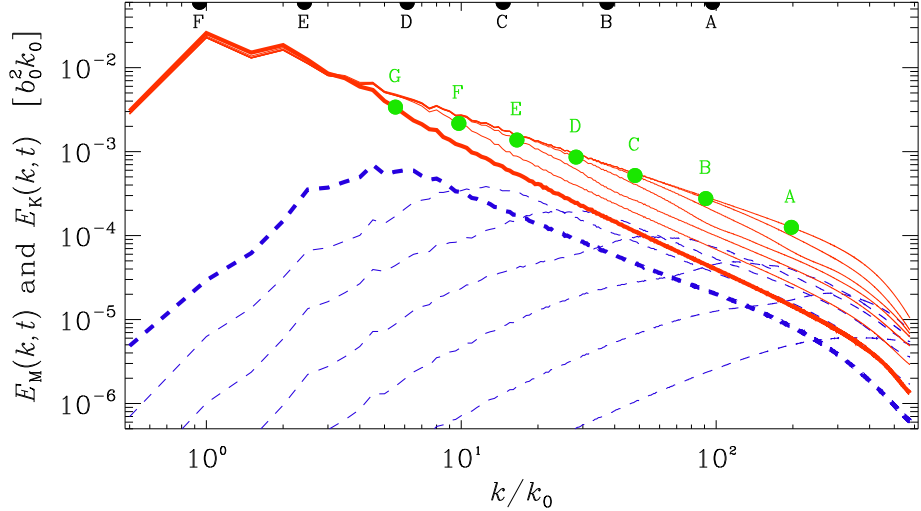}
\end{center}
\caption{
Magnetic (red) and kinetic (blue) energy spectra at early times after having
initialized the magnetic field with a spectrum of the form \Eq{eScInvSpec}.
The green symbols on the red lines denote the position of $k_\star(t)$,
as given in \Eq{kstar}, while the black symbols on the upper abscissa
denote the location of the horizon wavenumber $k_{\rm hor}(t)$.
The times for spectra indicated by the letters A--G are $c k_0 t=0.05$,
0.13, 0.34, 0.8, 2.1, 5.3, and 15 respectively.
} \label{pkt1152_rKH2304tnuk4km1b_sig1}
\end{figure}

Our considerations above have not addressed the time evolution of the
magnetic energy spectrum.
This is the purpose of the present section.
To see how the spectrum changes with time, we solve the
hydromagnetic equations for an isothermal relativistic
gas with pressure $p=\rho/3$ \cite{beo,Kahniashvili:2016bkp}
\begin{eqnarray}
{\partial\ln\rho\over\partial t}&=&-\frac{4}{3}\left(\nab\cdot\uu+\uu\cdot\nab\ln\rho\right)
+{1\over\rho}\left[\uu\cdot(\JJ\times\BB)+\eta\JJ^2\right], \\
{\partial\uu\over\partial t}&=&-\uu\cdot\nab\uu
+{\uu\over3}\left(\nab\cdot\uu+\uu\cdot\nab\ln\rho\right)
-{\uu\over\rho}\left[\uu\cdot(\JJ\times\BB)+\eta\JJ^2\right] \nonumber \\
&&-{1\over4}\nab\ln\rho
+{3\over4\rho}\JJ\times\BB+{2\over\rho}\nab\cdot\left(\rho\nu\SSSS\right), \\
{\partial\BB\over\partial t}&=&\nabla\times(\uu\times\BB-\eta\JJ),
\label{dAdt}
\end{eqnarray}
where ${\sf S}_{ij}=\half(u_{i,j}+u_{j,i})-\onethird\delta_{ij}\nab\cdot\uu$
is the rate-of-strain tensor,
$\nu$ is the viscosity, and $\eta$ is the magnetic diffusivity.
We consider a periodic domain with sidewalls $L$ and volume $L^3$,
so the smallest wavenumber is $k_{\min}=2\pi/L$.

In ref.~\cite{Kahniashvili:2016bkp}, we have already considered the case with
an initial $k^{-1}$ spectrum that extends all the way to $k=k_{\min}$,
i.e., with no $k^4$ subinertial range for small $k$.
However, as discussed above, somewhere beyond the event horizon, the
spectrum must effectively have a $k^4$ subrange.
To resolve the full wavenumber range, we consider a numerical resolution
of $2304^3$ mesh points, restricting ourselves to early times only.
We clearly see that at early times the spectrum does not change at
small wavenumbers.
As time goes on, smaller and smaller values of $k$ are affected by
the growing velocity field.
In \Fig{pkt1152_rKH2304tnuk4km1b_sig1} we show the temporal evolution
of the value of $k$ where the spectrum begins to depart from the initial $k^{-1}$ spectrum.
We see that this growth is well described by a turbulent-diffusive growth like
\EQ
k_\star(t)\approx\xiM(t)\,(\eta_{\rm turb}t)^{-1/2}
\approx(\urms\kM t/3)^{-1/2},
\label{kstar}
\EN
where $\eta_{\rm turb}\approx\urms/3\kM$ is an approximation to the
{\it turbulent} magnetic diffusivity \cite{SBS08}.
The ratio $3\eta_{\rm turb}/\eta$ is the magnetic Reynolds number,
which has values of around $20,000$ in Run~A and $500$ in Run~B.
The values of $k_\star(t)$, indicated by green dots, agree well with
the positions where the spectrum departs from the initial $k^{-1}$ subrange.
Note also that the horizon wavenumber $k_{\rm hor}(t)=(ct)^{-1}$
is always smaller than $k_\star(t)$.
Note that the $k^4$ subinertial range begins to appear within
the horizon for $c k_0 t>5$, corresponding to symbol F in
\Fig{pkt1152_rKH2304tnuk4km1b_sig1}.

\begin{figure}
\begin{center}
\includegraphics[width=.8\columnwidth]{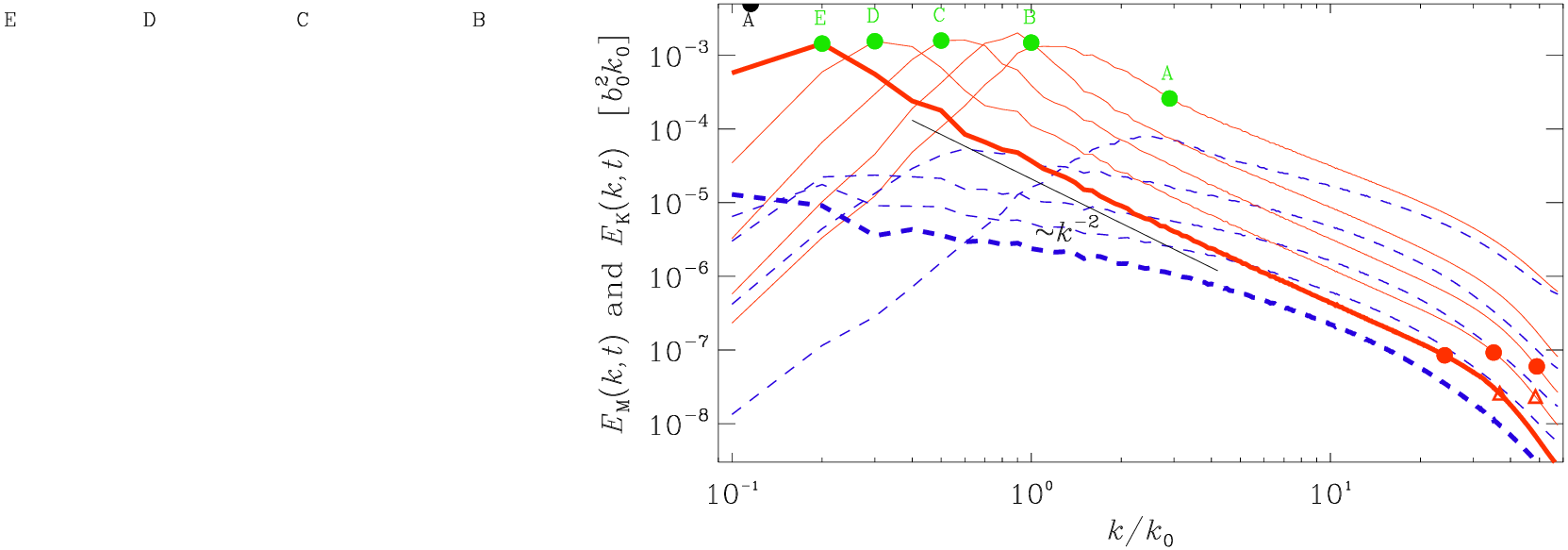}
\end{center}
\caption{
Similar to \Fig{pkt1152_rKH2304tnuk4km1b_sig1}, but for $k_0/k_{\min}=10$
and $1152^3$ meshpoints resolution.
The times are indicated by the letters A--E for $c k_0 t=9$,
70, 270, 900, and 2500.
The red triangles and red filled symbols on the respective magnetic energy
spectra denote the positions of $\kMD$ and $\kWT$, respectively, provided
they fall inside the plot range.
Note that $k_{\rm WT}$ is significantly smaller than $\kMD$,
for which all earlier times are still outside the plot range.
}\label{pkt1152_KH1152tnuk4km1a_sig1}
\end{figure}

The late time evolution is shown in \Fig{pkt1152_KH1152tnuk4km1a_sig1},
where we see that soon $\kM$ drops well below the initial value $k_0$.
This indicates the begin of the regular inverse cascade of
magnetic helicity.
The speed at which $\xiM$ increases is then governed by the usual
helicity evolution with $\xiM\propto t^q$, as can easily be derived
from dimensional arguments.

\section{Applications in cosmology}
\label{sec:Cosmology}

The Planck collaboration has recently published a comprehensive study
deriving upper limits for a primordial magnetic field based
on the measurements of CMB anisotropies and polarization;
see ref.~\cite{Ade:2015cva}.
They used the value of the  magnetic field smoothed over $1\,\mathrm{Mpc}$ to define upper bounds on the  magnetic
energy density and the magnetic helicity density; see
their equations~(2) and (13).

The smoothed fields used in~\cite{Ade:2015cva} are simply related to the ones defined in section~\ref{sec:normalization}\footnote{In their convention,
\begin{eqnarray}
{\bar B}_\lambda^2 &=& \int_0^\infty \frac{dk~k^2}{2\pi^2}
e^{-k^2\lambda^2} {\cal F}_{\rm M}(k),
\label{planckB}
\quad{\bar{\mathcal B}}_\lambda^2 = \lambda \int_0^\infty \frac{dk~k^3}{4\pi}
e^{-k^2\lambda^2} {\cal F}_{\rm H}(k). \label{planckH}
\end{eqnarray}}
 via $B_\lambda^2 = 2 {\bar B}_\lambda^2 $ and ${\bar{\mathcal B}}_\lambda^2 = {\mathcal H}_C$
In ref.~\cite{Ade:2015cva} the magnetic field spectrum is supposed to have a fixed
spectral shape in all regions with $k<k_2=k_D=$ the damping scale. They neglect the presence of a
turbulent regime for $k_I=k_1<k<k_D$ and characterize the symmetric magnetic power spectrum
by $E_{\rm M}(k) \propto k^{n_E}$ and the antisymmetric spectrum by $
H_{\rm M}(k) \propto k^{n_{\rm H}}$.
As we have shown above, such a description assumes
that $k_I \equiv k_D$ where $k_D$ is determined through the Alfv\'en
wave damping scale and is related to the amplitude of the magnetic field; see equation~(3) of ref.~\cite{Ade:2015cva}.
For a stochastic magnetic field that has a fixed spectral
slope up to the damping scale, $\EM(k) \propto k^{n_E}$ for $k\leq k_D$,
the Alfv\'en wave damping wavenumber depends on the rms magnetic field,
$B_{\rm rms} \equiv \sqrt{2\EEM}$ and the spectral index $n_E$ as
\cite{Kahniashvili:2010wm}, and is given by
\begin{equation}
\frac{k_D}{1~{\rm Mpc}^{-1}} = 1.4 \, \sqrt{
\frac{(2\pi)^{n_E+1} h}{\Gamma\big(\frac{n_E+3}{2}\big)}}
\left(\frac{10^{-7}~{\rm Gauss}}{B_{\rm rms}}\right)\,,
\end{equation}
where $h$ is the Hubble constant in units of 100 km s$^{-1}$ Mpc$^{-1}$.

As we have described above, we determine the magnetic damping scale
as a scale where the magnetic energy spectrum
decays exponentially, and the magnetic field power (in the short-wave
length scales) is negligible when compared to the long-wave length or
inertial regimes.
In \Fig{pkt1152_KH1152tnuk4km1a_sig1}, we indicate the positions of
$\kMD$ and $\kWT$.
In all cases, they are indeed much larger that the values
of $\kM$ and are therefore only important on very small
length scales.

More importantly, the magnetic field is evolving from the moment of
generation until recombination when the CMB photons decouple: this
evolution can be described as free decay in MHD
turbulence, which leads to a growth of the magnetic field correlation
length and the decay of the magnetic field amplitude, see \Sec{sec:Numerical} and \cite{Brandenburg:2016odr} for the decay laws.
Here we assume that the magnetic field has been generated with its maximal
\textit{comoving} strength of order of $10^{-6}$ Gauss.
This limit comes from BBN data, which limits the number or
relativistic species present at $T= T_{\rm BBN} \sim 0.1$MeV.
Since a magnetic field scales like a relativistic species, we can translate
this to a magnetic field amplitude, which is  $10^{-6}$ Gauss (comoving).

Magnetic fields generated at  the electroweak  cosmological phase transition are well below the current bounds (which are of the
order of $10^{-9}$Gauss) at  recombination, and thus these fields {\it will not leave any observable traces on the CMB} unless a mechanism that will significantly alter the magnetic field evolution via MHD which is discussed here (that seems to be unlikely).

The situation is somewhat more optimistic for magnetic fields generated during the QCD phase transition since the correlation length
is larger. But again, the amplitude and correlation scales of the obtained fields
are far too small to leave a detectable imprint on the CMB which requires $B_\lambda\gsim 10^{-9}$Gauss.

The bounds on the spectral index as obtained by the Planck collaboration
was due to their assumption of a flat prior distribution of the PMF. It has been shown
that such a bound vanishes when logarithmic priors are used \cite{Zucca:2016iur}.
This shows that a `blind trust' in Markov Chain Monte Carlo (MCMC)
results is problematic
and that the result depends on the choice of the prior.
It is probably fair to say that for a positive definite quantity, $A$, a flat prior in
$\log(A)$ is more appropriate than a flat prior in $A$.

In order to constrain the magnetic helicity, we need to observe parity odd
CMB spectra (such as EB) since the parity-even spectra (such as EE, BB, or
cross correlations) depend on both the symmetric and the helical parts
\cite{Caprini:2003vc,Kahniashvili:2005xe,Kunze:2011bp,Ballardini:2014jta}.
In other words, we need a measurement that depends solely on the helical component
in order to break the degeneracy between $\EM(k)$ and $\HM(k)$. This cannot be
provided by scalar quantities like the mean or rms energy densities, since it is
shown that they are independent of helicity,
and can be constrained through helicity-independent measurements such as the CMB Faraday rotation \cite{Campanelli:2004pm,Kahniashvili:2004gq,Kosowsky:2004zh}.

To constrain magnetic helicity, one has to consider parity odd CMB spectra
as outlined in~\cite{Caprini:2003vc,Kahniashvili:2014dfa,Kunze:2011bp}.
An upper bound on magnetic helicity can also be obtained via the
realizability condition if one can
limit independently the amplitude of the magnetic field (i.e., the
mean magnetic energy density, $\EEM$) and the correlation
length of the field. There are two independent ways of constraining the mean magnetic energy
density: (i) the CMB Faraday rotation measurement that is independent
of magnetic helicity \cite{Kosowsky:2004zh}; (ii) magnetic field
effects on the matter power spectrum, i.e., the limitation of the magnetic
field amplitude through LSS (in particular Ly-$\alpha$ statistics)
\cite{Kahniashvili:2012dy}.
The later gives stronger bounds of the order of nanoGauss.
The CMB fluctuations can be expressed in terms of the mean magnetic
energy density and the magnetic helicity density.
In fact, for maximally helical fields, a combination that determines
the strength of the parity-odd signal in the CMB, is
$\EEM\HHM$, see equation~(18) of Ref.~\cite{Kahniashvili:2014dfa}.
Thus, the upper bound of this quantity for  primordial magnetic fields
(independently of the magnetogenesis scenario)
is of the order of $10^{-18} \xiM$ Gauss$^2$.
For a causally generated magnetic field, the correlation length scale
must be less than the Hubble horizon at the moment of generation.
For  cosmological phase transitions, even accounting for
hydromagnetic turbulence decay, in the most optimistic scenario, the comoving
value of the correlation length is of the order of $30-50$ kpc, and thus
$\EEM\HHM$ is limited to be less than a few $10^{-20}$
Gauss$^2$ Mpc, while the CMB parity odd fluctuations might be sourced
if $\EEM\HHM$ is of the order of $10^{-17}$ Gauss$^2$ Gpc
which is 5--6 orders of magnitude larger than  what can be obtained
from  magnetic fields generated in a phase transition.
Therefore, we conclude that if magnetic helicity traces will be
detected on the CMB, it will be a direct indication that magnetic helicity
has originated in the inflationary epoch.

\section{Conclusions}
\label{sec:Concl}

In this paper we have addressed the main statistical properties of
helical magnetic fields, applying methods that are well established in
the theory of turbulence, mainly following statistical fluid dynamics
\`a la Monin and Yaglom \cite{monin-yaglom} and generalizing it to
the helical case.
We have also described in detail the different definitions of
helicity, such as magnetic, current, and kinetic helicity, and we have
made direct connections between them.
An important focus has been on the characteristic length scales
of the magnetic field such as the correlation and diffusion scales.
We have argued that the Alfv\'en damping scale used in earlier work
should be replaced by the proper diffusion length scales of the
turbulence where the scale-dependent Reynolds number is unity.

As expected, the energy density of the magnetic field does not depend
on the helical part of the correlation function (or the spectrum).
Also the rms value of the magnetic energy density is
independent of magnetic helicity, even though the four-point correlation
function of the magnetic field does contain information on the antisymmetric
part (quadratically) ~\cite{Caprini:2003vc,Kahniashvili:2014dfa,
Kunze:2011bp,Ballardini:2014jta}.

In addition to a theoretical study of the properties of nearly
scale-invariant helical magnetic fields and their evolution,
we have addressed its cosmological applications.
We have shown that, due to the magnetic decay since the moment of
generation until recombination,  causally generated magnetic field
cannot contribute to the CMB fluctuations at currently or near-future
observable levels.
Firstly, even in the most optimistic situation, the magnetic field strength
at generation is limited by the BBN bound being of the order of $10^{-6}$
Gauss.
Secondly, even if the correlation length of the magnetic field is of
the order of the Hubble scale at the moment of magnetogenesis and the
magnetic field experiences an inverse cascade, the correlation length at
recombination is much too small for such fields to leave an imprint on the CMB, see also Fig.~11 of Ref.~\cite{Brandenburg:2017neh}.

At this point {\it only} a nearly scale invariant spectrum possibly
generated during inflation might sustain the amplitude order of
$10^{-9}-10^{-10}$ Gauss with large enough correlation length scale and
with substantial magnetic helicity (bounded by the realizability
condition) can leave of any traces
on CMB maps which are accessible to present and near future observations.

\acknowledgments
It is our pleasure to thank Andrey Beresnyak, Leonardo Campanelli,
Gennady Chitov,  George Lavrelashvili, Arthur Kosowsky, Barnabas Poczos, Bharat Ratra,
Dmitri Ryabogin, Alexander G. Tevzadze, and Tanmay Vachaspati  for useful discussions.
We acknowledge partial support from the Swiss NSF SCOPES grant IZ7370-152581,
the Georgian Shota Rustaveli NSF grants FR/264/6-350/14 and FR/339/6-350/14, and the NSF
Astrophysics and Astronomy Grant Program grants AST1615940 \& AST1615100,
RD acknowledges support from the Swiss National Science Foundation.
T.K.\ and S.M.\ thank the Department of Physics at Geneva University where this
work has been initiated for hospitality.

\appendix

\section{Fourier transform of the magnetic two-point correlation function}
\label{FourierTwoPoint}

We present here the derivation of \Eq{spectral-density}.
The treatment follows \cite{Landau-Lifshitz}. We begin with the two-point correlation function
\begin{equation}
{\mathcal B}_{ij} ({\bf r}_1, {\bf r}_2) = \langle B_i({\bf r}_1)B_j({\bf r}_2)\rangle.
\end{equation}
To compute its Fourier transform, we first write
\begin{equation}
{\mathcal B}_{ij} ({\bf r}_1, {\bf r}_2) =
\frac{1}{(2\pi)^6}\int d^3 \mathbf k\,\int d^3\mathbf k' \, \langle
B_i^\star(\mathbf k)B_j(\mathbf k')\rangle
e^{-i(\mathbf k \cdot \mathbf r_1 - \mathbf k'\mathbf r_2)}.
\label{Bij-fourier}
\end{equation}
Statistical homogeneity implies that $B_{ij}({\bf r}_1, {\bf r}_2)$ is a function of
$\mathbf r=\mathbf r_2-\mathbf r_1$ only, so we must have
\begin{equation}
\langle B_i^\star(\mathbf k)B_j(\mathbf k')\rangle = (2\pi)^3\delta^{(3)}(\mathbf k-\mathbf k'){\mathcal F}_{ij}(\mathbf k),
\end{equation}
for some function ${\mathcal F}_{ij}(\mathbf k)$. Then,
\begin{equation}
{\mathcal B}_{ij}(\mathbf r) =
\frac{1}{(2\pi)^3}\int d^3\mathbf k \,e^{-i\mathbf k\cdot\mathbf r}{\mathcal F}_{ij}(\mathbf k).
\end{equation}

\section{Root-mean-square magnetic energy density}
\label{a:B}

The purpose of this appendix is to present the detailed
derivation of \Eq{EEMrms}.
We compute the rms magnetic energy density given as,
\begin{equation}
\EEM^{\rm rms} = \left(\frac{1}{(2\pi)^3} \int  d^3 {\bf k} \,{\mathcal R}_{\rm M} ({\bf k}) \right)^{1/2} \,,
\label{rho22}
\end{equation}
where ${\mathcal R}_{\rm M}({\bf k})$ is the power spectrum defined through
\begin{equation}
\langle \rho_{\rm M}^\star({\bf k}) \rho_{\rm M}({\bf k^\prime}) \rangle = (2\pi)^3 \delta^{(3)}
({\bf k} - {\bf k'}) {\mathcal R}_{\rm M}({\bf k}).
\label{rho2}
\end{equation}
A short calculation using Wick's theorem gives
\begin{equation}
 {\mathcal R}_{\rm M}({\bf k}) = \frac{\delta^{(3)}({\bf k})}{4(2\pi)^6}\left(\int d^3\mathbf{p} {\cal F}_{ii}({\bf p})\right)^2 + \frac{1}{2(2\pi)^6}\int d^3\mathbf{p} {\cal F}_{ij}({\bf p}){\cal F}_{ij}({\bf k}-{\bf p}) \,.
\end{equation}
In the next step we use expression (\ref{correlfourier}) for ${\mathcal F}_{ij}({\bf k})$. With this we obtain
\begin{eqnarray}
{\mathcal R}_{\rm M}({\bf k}) &= &
{\cal I}_1 ({\bf k}) + {\cal I}_2({\bf k}) + {\cal I}_3({\bf k})= \frac{\delta^{(3)}({\bf k})}{(2\pi)^6}\left(\int d^3\mathbf{p} {\cal F}_{\rm N}({ p})\right)^2
\nonumber \\
&+&
\frac{1}{(2\pi)^6}\int d^3\mathbf{p} {\cal F}_{\rm N}({p}){\cal F}_{\rm N}(|{\bf k}-{\bf p}|)(1+\mu^2)
+
\frac{1}{(2\pi)^6}\int d^3\mathbf{p} {\cal F}_{\rm H}({ p}){\cal F}_{\rm H}(|{\bf k}-{\bf p}|)\mu \, ,
\label{Rk}
\end{eqnarray}
where $\mu={\hat {\bf p}} \cdot {\widehat{ ({\bf k}-{\bf p})}}$

To compute the rms of the magnetic field energy density we
use Eq.~(\ref{rho22}), and we obtain
\begin{equation}
(\EEM^{\rm rms})^2 = {\cal I}_1 + {\cal I}_2 + {\cal I}_3
\end{equation}
with

\begin{eqnarray}
{\cal I}_{1} &=& \frac{1}{(2\pi)^6}\int d^3 {\bf k} \, \delta^{(3)}({\bf k})
\left(\int d^3\mathbf{p} {\cal F}_{ii}({\bf p})\right)^2
= \frac{1}{(2\pi)^6} \left(\int d^3 {\bf p} \,  {\cal F}_{\rm N}(p)\right)^2 \, ,
\label{A1}
\\
{\cal I}_2 &=&\frac{1}{(2\pi)^6} \int d^3\mathbf k\, \int d^3 {\bf p}\, {\cal F}_{\rm N}(p) {\cal F}_{\rm N}(|\mathbf k-\mathbf p|)\, (1+\mu^2) \, ,
\label{A12}
\\
{\cal I}_{3} & = &\frac{4}{(2\pi)^6} \int d^3\mathbf k\, \int d^3 {\bf p}\, {\cal F}_{\rm H}(p) {\cal F}_{\rm H}(|\mathbf k-\mathbf p|) \, \mu \, .
\end{eqnarray}
To proceed we use  the variables transform
 $\mathbf q = \mathbf k-\mathbf p$, so $d^3\mathbf q = d^3\mathbf k$.

The integral ${\cal I}_{1}$ is simply
\begin{eqnarray}
{\cal I}_{1} = \frac{1}{3}\EEM^2.
\label{SF}
\end{eqnarray}
Under the exchange of variables the integral ${\cal I}_2$ becomes
\begin{equation}
{\cal I}_2=\int d^3\mathbf q\,\int d^3\mathbf p\,  {\cal F}_{\rm N}(p)  {\cal F}_{\rm N}(q) \Big[ 1+
\big({\hat {\bf p}} \cdot {\hat {\bf q}} \big)^2 \Big]\,.
\end{equation}
Now, let us write
\begin{equation}
\mu= ({\hat {\bf p}} \cdot {\hat {\bf q}})= \frac{4\pi}{3} \sum_{m=-1}^{1}
Y^\star_{1m}({\bf {\hat p}}) Y^\star_{1m}({\bf {\hat q}}),
\label{pq}
\end{equation}
where $Y_{lm}({\hat{ {\bf k}}})$ are the spherical harmonics.
Thus,
\begin{equation}
\mu^2= \Big( \frac{4\pi}{3} \Big)^2  \sum_{m=-1}^{1} \sum_{m'=-1}^{1} \Big[
Y^\star_{1m} ({\hat{ {\bf p}}}) Y_{1m'} ({\hat{ {\bf p}}}) Y_{1m} ({\hat{ {\bf q}}}) Y^\star_{1m'} ({\hat{ {\bf q}}})\Big],
\end{equation}
and we use
\begin{equation}
\int_{\Omega_k} d^2\hat{\mathbf{n}}\,Y^\star_{1m} ({\hat{\bf k}}) Y_{1m'} ({\hat{ {\bf k}}}) = \delta_{mm'} \,,
\end{equation}
so that
\begin{equation}
\int d^2\hat{ {\bf p}}\int d^2\hat{ {\bf q}} \Big[ 1+
\big({\hat {\bf p}} \cdot {\hat {\bf q}} \big)^2 \Big] = (4\pi)^2\left[1 +\frac{1}{3}\right]\,, ~\mbox{ and }~  {\cal I}_2 = \frac{4}{3}\EEM^2.
\end{equation}
Finally,
to compute ${\cal I}_3$,
as above, we use exchange of variables and get, using Eq.~(\ref{pq}),
\begin{equation}
{\cal I}_{3}=\frac{4}{(2\pi)^6} \int d^3 {\bf q} \, \int d^3 {\bf p} \,
 {\cal F}_{\rm H}(p)  {\cal F}_{\rm H} (q) \sum_{m=-1}^1 Y^\star_{lm} ({\hat{ {\bf p}}}) Y_{lm} ({\hat{ {\bf q}}}) =0,
\label{AF}
\end{equation}
where we have used $\int d^2 {\bf \hat{\bf n}} Y_{1m}(\hat{\bf n})=0$.
Collecting all the terms gives $\EEM^{\rm rms}=\sqrt{\frac{5}{3}}\EEM$.

We have seen that ${\cal I}_3$ given by the double
integral over $\mathbf{p}$ and $\mathbf{k}$ vanishes.
However, the angular $k$ integral
of eq.~(\ref{A12}) is finite for small $k$, as seen in the bottom left panel
of figure~1 of ref.~\cite{Ballardini:2014jta}.
This angular integral is defined as
\begin{equation}\label{AngI3}
{\cal J}_3(k)=\int_{-\infty}^\infty d^3\bm{p} \int_{4\pi} k^2 d\Omega_k\,
{\cal F}_{\rm H}(p) {\cal F}_{\rm H}(q) \; \bm{\hat{p}} \cdot \bm{\hat{q}}
\end{equation}
and it satisfies ${\cal I}_3=\frac{4}{(2\pi)^6} \int_0^\infty dk\,k^2\,{\cal J}_3(k)=0$.

\begin{figure*}[h!]\begin{center}
\includegraphics[width=.7\textwidth]{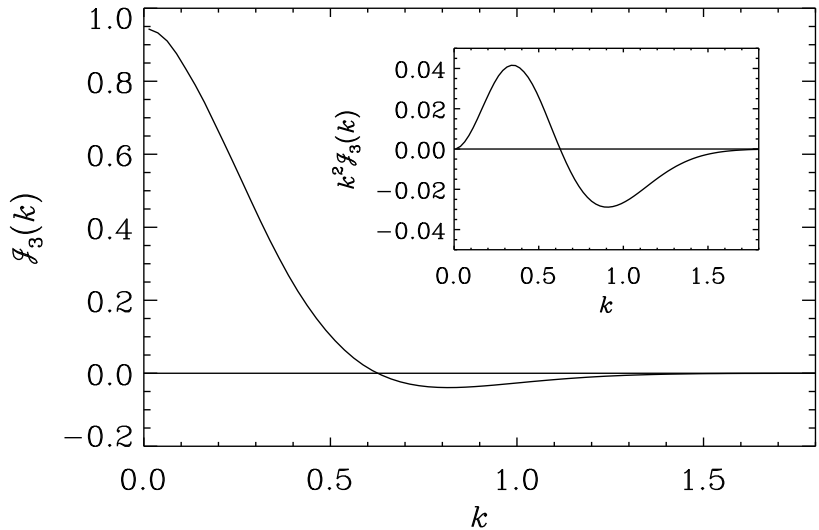}
\end{center}\caption[]{
Result for ${\cal J}_3(k)$.
The inset shows $k^2{\cal J}_3(k)$.
The areas underneath the positive and negative parts are equal, so
$\int dk\,k^2{\cal J}_3(k)=0$.
}\label{J3result}\end{figure*}

To gain some insight into the functional form of ${\cal J}_3(k)$,
we adopt as an example a Gaussian for ${\cal F}_{\rm H}(p)$, i.e.,
\begin{equation}
{\cal F}_{\rm H}(p)=\exp(-p^2/2p_0^2),
\end{equation}
with $p_0=0.3$ and compute ${\cal J}_3(k)$ numerically.
The result is shown in figure~\ref{J3result}.
We see that, although $\int_0^\infty dk\,k^2{\cal J}_3(k)=0$, the integral
$\int_0^{k_{\max}} dk\,k^2{\cal J}_3(k)$ does not vanish for sufficiently
small values of $k_{\max}$.

In the early universe, due to the high conductivity, the Reynolds numbers are very high, and it
is reasonably expected that $k_{\rm max}\gg k_I$,\footnote{
For Kolmogorov spectra, the Reynolds number is given by
${\rm Re} = ( k_D/k_I )^{4/3}$ thus $k_D \gg k_I$,}
but we see that the integral converges to zero
if $k_{\rm max}/k_I\gtrsim\mathcal{O}(1)$.
There might exist some integral measurable quantity
determined by the intermediate value $k_q$, such that
$\tilde{\cal I}_3=\int_0^{k_{q}} dk\,k^2\,{\cal J}_3(k)$ is nonvanishing.
In such a case, care must be taken to interpret those quantities, since the dependence
on helicity in that case could be affected by measurement details.
The sign of magnetic helicity, however, cannot enter such a dependence.

\end{document}